%% file: ms.tex
\documentclass[12pt,preprint]{aastex}

\received{}
\accepted{}
\journalid{}{}
\articleid{}{}

\slugcomment{Astronomical Journal, submitted}

\shortauthors{Landolt}
\shorttitle{$UBVRI$ Equatorial Standard Stars}

\begin{document}

\title{UBVRI PHOTOMETRIC STANDARD STARS AROUND THE CELESTIAL EQUATOR:  UPDATES AND ADDITIONS}

\author{Arlo U.~Landolt\altaffilmark{1}}
\affil{Department of Physics \& Astronomy, Louisiana State University, Baton Rouge, LA 70803-4001}
\email{landolt@phys.lsu.edu}

\altaffiltext{1}{Visiting Astronomer, Cerro Tololo Inter-American Observatory, National Optical 
Astronomical Observatory, which is operated by the Association of Universities for Research in 
Astronomy, Inc., under contract with the National Science Foundation.}

\begin{abstract}
\label{sec:abstract}

New broad-band $UBVRI$ photoelectric observations on the Johnson-Kron-Cousins photometric system 
have been made of 202 stars around the sky, and centered at the celestial equator.  These stars 
constitute both an update of and additions to a previously published list of equatorial photometric 
standard stars.  The list is capable of providing, for both celestial hemispheres, an internally 
consistent homogeneous broadband standard photometric system around the sky.  When these new 
measurements are included with those published in \citet{Landolt1992}, the entire list of standard 
stars in this paper encompasses the magnitude range $8.90<V<16.30$, and the color index range 
$-0.35<(B-V)<+2.30$.

\end{abstract}


\section{Introduction}
\label{sec:introduction}

Over the past thirty five years, the author has published photometric standard star papers 
\citep{Landolt1973, Landolt1983, Landolt1992, Landolt2007a, LandoltUomoto2007} wherein the $UBV$ 
magnitudes and color indices always have been based on the defining paper of \citet{Johnson1963}, 
and the $RI$ photometry has been based on the work of \citet{Cousins1976}.  A summary of the 
author's observing and analysis procedures may be found in \citet{Landolt2007b}.

The current paper essentially is an update and extension of of the photometry which appeared in 
\citet{Landolt1992}.

\section{Observations}
\label{sec:observations}

New broad-band $UBVRI$ photoelectric observations on the Johnson-Kron-Cousins photometric system have 
been made of 202 stars around the sky, and centered at the celestial equator.  One hundred thirty two 
(132) of these 202 stars update magnitudes and color indices in \citet{Landolt1992}.  The other 70 
stars are completely new additions to that list of standard stars.

The new photometric results in this paper represent both an update of and addition to the list of 
photometric standard stars in \citet{Landolt1992}.  Most of these photometric standard stars inhabit 
a band of less than five degrees width, centered on the celestial equator, around the celestial	
sphere.

Stars whose photometry was upgraded came primarily from \citet{Landolt1992}.  However, several 
brighter stars taken from \citet{Landolt1983} also were observed and included.  New photometric 
sequences have been established, anchored by blue stars which already were standard stars themselves 
\citep{Landolt1992}.  Examples of such stars are Feige~24 and G~93-48.

All of the new data in this paper were taken at the CTIO 1.5-m telescope together with GaAs 
photomultipliers.  The author always has tried to maintain the dictum \citep{Landolt2007b} that 
potential standard star data only should be obtained with one detector.  However, the	
photomultipliers did not cooperate; four different GaAs photo-multipliers were used during this 
observational program, albeit the majority of the data were collected with one photomultiplier.

A total of 224 nights were assigned over a nine year time interval.  Useful data were obtained on 
150 of the assigned nights, implying that 66.7\% of the nights were photometric.  From another 
viewpoint, a total of $2,206$ night-time hours were assigned.  Useful data were obtained during 
$1,192$ hours, indicating that 54.1\% of the assigned hours were photometric.  It should be noted 
that heretofore, the author always has quoted the number of nights which were photometric, which is 
the method used by most observers.  There is a difference.

The initiation of this program encountered problems with successive photomultipliers.  The 
photomultipliers with which data were obtained, including Universal Time (UT) dates were: a standby 
RCA31034, serial no. C20453 in cold box 58 in the interval 14 June 1993 through 17 June 1993; a 
Hamamatsu R943-02, serial no. EA 4267 in cold box 50 in the interval 20 May 1994 to 24 July 1995; the 
LSU C31034A-02 in cold box 60 in the interval 13 March 1996 through 8 May 1998; and a RCA 31034A-02 
in KPNO cold box 53 in the interval 7 July 1998 through 11 December 2001. Forty-three (43) nights of 
useful data were taken with the first three photomultipliers. One hundred sixteen (116) nights of 
useful data were taken with the last identified photomultiplier.

Each photomultiplier had a history.  The photomultiplier in cold box 58 was meant to be, and was used 
as a standby, as a search was made for a stable photomultiplier.  Since its sensitivity was low, it 
only was used for one observing run, and contributed useful data on four nights of observing.  After 
roughly one year of use, the Hamamatsu developed an instability. It was replaced by the LSU 
C31034A-02, so-called because the author's institution paid for it.  That photomultiplier, 
C31034A-02, was a successor to RCA 31034A-02 models, and was manufactured by Burle Industries, which 
had purchased the old RCA tubes' division.  The ``LSU" photomultiplier also developed a problem, 
never successfully repeated or diagnosed in the laboratory.  In order to be as confident as possible 
in data being acquired, a switch was made to the fourth photomultiplier, a RCA 31034A-02 in KPNO cold 
box 53.  That tube became available with the closure of photoelectric photometry programs at Kitt 
Peak.

All photomultipliers were operated at $-1600V$.  None of the photomultipliers' sensitivity functions 
were available.  Data for the RCA brand of photomultiplier has been tabulated in \citet{Landolt1992} 
in Appendix B, Table 11, and are illustrated in Figures 52-54.

The filter set used throughout the observational program was CTIO's $UBVRI$ filter set \#3.  
Information describing the composition of that filter set may be found in \citet[][Table 
III]{Landolt1983}.  The transmission characteristics of those filters are tabulated in 
\citet[][Appendix B]{Landolt1992}.

Between 20 and 25 $UBVRI$ standard stars, as defined by \citet{Landolt1992}, were observed each night 
together with the program stars.  A night's observations began and ended with a group of four or five 
standard stars.  Similar groups were observed periodically throughout the night.  Each of these 
groups contained stars closely spaced on the sky, and possessing as wide a color range as possible.  
A more complete outline of the author's observing philosophy has been given in \citet{Landolt2007b}.

A complete data set for a star consisted of a series of measures: $VBURIIRUBV$.  A $14.0\arcsec$ 
diaphragm was used throughout the observing program.  The integration or counting time depended upon 
the faintness of a particular star.  The counting time never was less than ten seconds per filter, 
and was as long as $60\,$s for the faintest stars.  Data reduction procedures followed the precepts 
outlined by \citet{SchulteCrawford1961} and by \citet{Landolt2007b}.  It should be noted that the 
author always has reduced the $(V-I)$ color index independently.  That is, the author's $(V-I)$ color 
index values are not a direct combination of the $(V-R)$ and $(R-I)$ color indices.

Extinction coefficients were calculated from three or four standard stars possessing a range in color 
index that were followed from near the meridian over to an air mass of 2.1, or so.  Each night's data 
were reduced using the primary extinction coefficients derived from that night, whenever possible.  
Average secondary extinction coefficients for a given run were used.  Examples of the range in 
extinction coefficients which an observer in fact encounters have been tabulated in 
\citet{Landolt2007b}.  Such tabulations should remind any observer of the perils in using mean 
extinction coefficients.

The final computer printout for each night's reductions contained the magnitude and color indices for 
each of the standard stars.  Since the time of observation was recorded for each measurement, it was 
possible to plot the residuals in the $V$ magnitude and in the different color indices for each 
standard star against Universal Time for a given night.  These plots permitted small corrections to 
be made to all program star measures.  The corrections usually were less than a few hundredths of a 
magnitude.  Such corrections took into account small changes in both atmospheric and instrumental 
conditions that occurred during the course of a night's observations.

\section{Discussion}
\label{sec:discussion}

The magnitudes and color indices for 132 stars which appeared in \citet{Landolt1992} have been 
improved in accuracy via additional measurements.  Further, new stars were added to some fields 
(e.g., Feige~24 and G~93-48) thereby creating photometric sequences with a more broad range in color 
index.  There are 70 new additions to the photometric standard stars in this new edition.

The accuracy of the magnitude and color index transformations was checked via a comparison of the 
magnitudes and color indices of the stars from \citet{Landolt1992} that were used as standards in 
this paper, with the magnitudes and color indices of these same stars obtained during this current 
project.  The comparisons, the delta quantities, were in the sense of data from this project minus 
the corresponding magnitudes and color indices from \citet{Landolt1992}.  The plots of these delta 
quantities as a function of the \citet{Landolt1992} standard stars' magnitudes and color indices are 
not illustrated in this paper, since overall the general appearance of the plots is the same as in 
the author's previous papers \citep[for example, see Figs. 1-6 in][]{Landolt2007a}.  And, since four 
photomultipliers were used, this means a savings of 24 plots.

As in the past, an inspection of each plot, the delta quantities on the ordinate, and the color 
indices on the abscissae, show the presence of nonlinearities in the transformation process. 
Inspection of each figure allowed the nonlinear ``break points" to be chosen.  The break points for 
these data are the same as those found in \citet{Landolt2007a}.  They are indicated in Table 
\ref{tab:table1} along with the appropriate nonlinear transformation relations, which were derived by 
least-squares fitting from the plots of the delta quantities as a function of the \citet{Landolt1992} 
standard stars' magnitudes and color indices.

The nonlinear transformation relations had the form in which a subscript ``c" indicates ``catalogue" 
and subscript ``obs" indicates ``observed."  As an example, the first line in Table \ref{tab:table1} 
has the form:

\begin{eqnarray}
(B-V)_{c}&=&-0.00212 + 1.03758(B-V)_{obs}~~~~~~~~~~~~~~~~~~~~~~~~(B-V)<+0.1, \nonumber \\
& & \pm 0.00284 \pm 0.01959 \nonumber \\
\end{eqnarray}

\noindent{i.e., each color index is corrected as a function of itself.  Note that the correction 
to the $V$ magnitude has to be made after the $(B-V)$ non-linear transformation correction has 
been done:}

\begin{eqnarray}
V_{c}&=&-0.00331 - 0.01849(B-V)_{c} + V_{obs}~~~~~~~~~~~~~~~~~~~~(B-V)<+0.1, \nonumber \\
& & \pm 0.00629 \pm 0.04330 \nonumber \\
\end{eqnarray}

The nonlinear transformation relations in Table \ref{tab:table1} were applied to the recovered 
magnitudes and color indices of the standard stars used in this project.  The data then were on the 
broadband $UBVRI$ photometric system, as defined by the standard stars in \citet{Landolt1992}. This 
process was done separately for data collected with each photomultiplier.  Next, the standard star 
magnitudes and color indices, now corrected for nonlinear transformation, were compared once again to 
the published values in the sense of corrected values minus published magnitudes and color indices.  
The fact that the nonlinear effects were corrected successfully could be illustrated with an 
additional 24 plots, not shown, again since their appearance is as in previous papers \citep[for 
example, see Figs. 7-12 in][]{Landolt2007a}. Therefore, the data in this paper obtained with each 
photomultiplier have been transformed to the photometric system defined in \citet{Landolt1992}.

All the data obtained with the four photomultipliers, and corrected with nonlinear transformation 
relations, plus previous standard star data from \citet{Landolt1992} for stars in common, were 
combined to produce Table \ref{tab:table2}, the final table of magnitudes and color indices.  The 
data were weighted by the number of observations and inversely as the square of the mean error of a 
single observation via the relations \citep{Barford1967},

\begin{equation}
\rm{weighted~result} = \left( \frac{\omega_{1} X_{1} / s^{2}_{1}  +  \omega_{2} X_{2} / s^{2}_{2}}{\omega_{1} / s^{2}_{1} + \omega_{2} / s^{2}_{2}} \right),
\end{equation}

\begin{equation}
\rm{weighted~error} = \left( \frac{\omega_{1} + \omega_{2}}{\omega_{1}/s^{2}_{1} + \omega_{2}/s^{2}_{2}} \right) ^{1/2},
\end{equation}

\noindent{where $\omega_{i}$ is the number of nights, $X_{i}$ is the magnitude or color index, and 
$s_{i}$ is the mean error of a single observation.}

There are 595 photometric standard stars in Table \ref{tab:table2}.  Of these 595 stars, 413 have 
five or more observations each.  These 413 stars are the most robust to use for standardization 
purposes.  Further, when choosing stars in Table \ref{tab:table2} to use as standard stars, whenever 
possible, the observer should choose stars with many individual observations, as well as those with 
small errors.

Finding charts are provided via Figures \ref{fig:figure1} - \ref{fig:figure136}.  The coordinates 
in Table \ref{tab:table2} were taken from the UCAC2 catalogue \citep{Zacharias2004} when possible.  
Positions for stars not in the UCAC2 catalogue were taken from the 2MASS Point Source Catalogue, 
whose coordinates came from the Two Micron All Sky Survey \citep[2MASS;][]{Skrutskie2006}.

Columns (4)-(9) in Table \ref{tab:table2} give the final magnitude and color indices in the $UBVRI$ 
photometric system as defined by \citet{Landolt1992}.  Column (10) indicates the number of times $n$ 
that each star was observed.  Column (11) gives the number of nights $m$ that each star was observed.  
The numbers in columns (4)-(9) are mean magnitudes and color indices.  Hence, the errors tabulated in 
columns (12)-(17) are mean errors of the mean magnitude and color indices \citep[see][p. 
450]{Landolt1983}.

The 595 stars in Table \ref{tab:table2}, on average, have been observed 24 times each on 17 different 
nights.  An error analysis for all the stars in Table \ref{tab:table2} is given in Table 
\ref{tab:table3}.  The numerical size of the average mean error of a single observation of a $V$ 
magnitude or a color index for the 595 stars in Table \ref{tab:table2} is given in column (2) of 
Table 3.  Column (3) shows the average mean error of the mean observed magnitude or color index.  
Errors in column (2) for a single observation are as large as they are for $(U-B)$, $(R-I)$, and 
$(V-I)$, since red stars are faint in $U$ and blue stars are faint in $I$.

Although accurate coordinates are necessary for individual stars in many circumstances, modern area 
detectors need knowledge of the coordinate center of a photometric sequence.  Table \ref{tab:table4} 
provides the coordinate centers for the $UBVRI$ photometric sequences listed in Table 
\ref{tab:table2}.  The field name is based on a Selected Area, or on a star, usually a blue star, 
chosen from the literature years ago on the basis that it was blue.  An exception is the T~Phe field. 
T~Phe is a Mira	variable (= HD~$2725$ = CD$-47~131$ = CPD$-47~50$ = GSC~$08024-01000$) whose 
photometric sequence was included in \citet{Landolt1992} just because the author had completed the 
sequence in an area of the sky (near the Magellanic Clouds) where another photometric sequence would 
be of use.  The current T~Phe sequence has been enlarged from the sequence initially published in 
\citet{Landolt1992}; it also appears in \citet{Landolt2007a}.

The author has gone another step.  Most of the Selected Area fields are 60 or 70 arc minutes on a 
side.  Many detectors, however, encompass much smaller fields of view.  Hence, where it makes sense to 
do so, the Selected Area fields have been subdivided into smaller fields, usually 10 arc minutes on a 
side, positioned to enclose as many standard stars as possible into the 10x10 arc minute field of 
view.  The smaller field within a Selected Area has been named a sub-field (SF), e.g., SA92 SF1 is 
sub-field 1 within Selected Area 92.  The coordinates of the center of each sub-field are given in 
Table \ref{tab:table4}.

The range in magnitude and color index for each field or sub-field is given in Table \ref{tab:table4}.  
The last column in Table \ref{tab:table4} lists the stars within each standard star field or 
sub-field.  Those stars with fewer than five measures each appear in italics in the last column of 
Table \ref{tab:table4}.  It is hoped that this information will aid in identifying which stars are to 
be used for a given sub-field during data reductions.

Some observers may find it to their advantage to redefine the centers for the Selected Area fields 
to better fit their detectors' needs.

The magnitude distribution of the stars in Table \ref{tab:table2} with five or more observations 
each is plotted in Figure \ref{fig:figure137} in $0.25$ $V$~mag bins.  Figure \ref{fig:figure138} 
shows the $(B-V)$ color index range in $0.1$~mag bins for these same stars.

Figures \ref{fig:figure139} - \ref{fig:figure145} have been plotted using data for the stars in 
Table \ref{tab:table2} with five or more observations each.  These figures show the mean error of 
a mean magnitude or color index, plotted as a function of magnitude or color index.

Figures \ref{fig:figure146} and \ref{fig:figure147} illustrate the $[(U-B), (B-V)]$ and $[(V-R), 
(R-I)]$ color-color plots for all stars with five or more measurements each.  The two stars in 
Figure \ref{fig:figure146} which fall to the upper right of the main sequence are SA98-L5 
[$(U-B)=-0.100$; $(B-V)=+1.900$] and SA110-273 [$(U-B)=+1.000$; $(B-V)=+2.527$].  The associated 
photometric errors are large, especially for SA98-L5, and hence were not plotted in Figures 
\ref{fig:figure139} - \ref{fig:figure145}.  Similarly, the $(U-B)$ errors for the stars SA110-157, 
SA110-315 and SA110-273, large due to the small flux in the $U$-band, were not plotted in the 
relevant figures.

Cross-identifications are provided in Table \ref{tab:table5}.  On occasion the very best coordinates 
and proper motion information is needed for standard stars.  Hence, Table \ref{tab:table5} presents 
the most recent coordinates and proper motions for the new standard stars in Table \ref{tab:table2}.  
All coordinates are for the epoch J2000.0.  The 2MASS Point Source Catalogue (PSC) positions come 
from The Two Micron All Sky Survey \citep[2MASS;][]{Skrutskie2006}.  The UCAC2 positions come from 
The Second USNO CCD Astrograph Catalogue \citep{Zacharias2004}.  Although attempts have been made to 
eliminate them, users should be aware that there may be erroneous or spurious proper motions in Table 
\ref{tab:table5} \citep[see][]{Levine2005}.

Spectral types for many of the Selected Area stars, classified by Drilling, may be found in 
\citet{DrillingLandolt1979}.  In addition, there exists an all-sky comprehensive spectral 
classification file prepared by \citet{Skiff2008}.  This file may be queried at the Simbad 
Strasbourgh Vizier site\footnote{see http://webviz.u-strasbg.fr/viz-bin/VizieR} by entering ``B/mk" 
in the catalogue search area.

\section{Comments on Individual Stars}
\label{sec:starcomments}

A number of the stars in Table \ref{tab:table2} are of interest for several reasons.  Therefore, in 
this section, additional references are given for selected stars.  Most of these stars have 
additional identifications.  A description follows next of the stellar nomenclature which the reader 
will encounter.  These different star naming systems and catalogues are given in alphabetical order.

\noindent{$\bf{BD}$: The Bonner Durchmusterung is a multiple volume star catalogue published between 
1859-62 by \citet{Argelander1859, Argelander1861, Argelander1862}.  It is a visual survey of stars in 
declination zones from $+90$ to $-01$ degrees.  The Sudliche Bonner Durchmusterung 
\citep{Schoenfeld1886} covers the declination range from $-02$ to $-22$ degrees.}

\noindent{$\bf{BPM}$: The Bruce Proper Motion survey was published by \citet{Luyten1963a} in two 
volumes.}

\noindent{$\bf{BPS}$: A catalogue of candidate field horizontal-branch stars was published by 
\citet{Beers1988}.}

\noindent{$\bf{Ci}$: High proper motion stars published in the Publications of the Cincinnati 
Observatory.  The prefix Ci~18 means the star was published in Publication No. 18; the prefix Ci~20 
means the star appeared in Publication No. 20.}

\noindent{$\bf{CD}$: The Cordoba Durchmusterung is a visual survey of southern stars in the 
declination zones $-22$ to $-90$ degrees, carried out by \citet{Thome1892, Thome1894, Thome1900, 
Thome1914, Thome1932}, and published in the Resultados del Observatorio Nacional Argentino as an 
extension to the Bonner Durchmusterung (BD) Catalogue.  The CD contains 613,959 records for stars 
brighter than $10.0$ magnitude.}

\noindent{$\bf{EGGR}$: An acronym made from the first initials of EGgen (Olin J.~Eggen) and 
GReenstein (Jesse L.~Greenstein).  Earlier acronyms for the same stars were EG and GR.  For 
additional details, and a summary of the papers wherein these stars were published, go into Simbad, 
look up Feige~22, for example, and click on EGGR~19.}

\noindent{$\bf{FASTT}$: Stars in a list emanating from the Flagstaff Astronometric Scanning Transit 
Telescope (FASTT); see a paper by \citet{HendenStone1998}.}

\noindent{$\bf{Feige~(F)~stars}$: A search for faint blue stars \citep{Feige1958, Feige1959}.}

\noindent{$\bf{GCRV}$: General Catalogue of Stellar Radial Velocities \citep{Wilson1953}.}

\noindent{$\bf{G,~GD,~and~GR}$: Proper motion stars published by Giclas and colleagues in the Lowell 
Observatory Bulletins \citep{Giclas1971}: Northern Hemisphere - the G numbered stars; 
\citet{Giclas1978}: Southern Hemisphere Catalogue.  \citet{Giclas1980}: Summary catalogue of the GD 
and GR stars.  This latter group contained very blue (GD) or very red (GR) stars of little or no 
proper motion.}

\noindent{$\bf{GJ}$: The nomenclature GJ pertains to stars in the \citet{GlieseJahreiss1979} catalogue 
of nearby stars.}

\noindent{$\bf{GSC}$: The Hubble Space Telescope Guide Star Catalogue (GSC) acronym first appeared in 
\citet{Lasker1990}.}

\noindent{$\bf{HD}$: The Henry Draper catalogue was published in the Annals of the Harvard College 
Observatory, vols. 91-99 in the time interval 1918-1924 \citep{CannonPickering1918a, 
CannonPickering1918b, CannonPickering1919a, CannonPickering1919b, CannonPickering1920, 
CannonPickering1921, CannonPickering1922, CannonPickering1923, CannonPickering1924}.  It contains 
coordinates, magnitudes and spectral types for 225,300 stars across the celestial sphere.  These 
volumes form the basis for astronomical spectral classification, and are the source of the OBAFGKM 
terminology used in remembering the sequence of spectral classes.}

\noindent{$\bf{IDS}$: The Index Catalogue of Visual Double stars \citep{Jeffers1963} is a catalogue 
of visual binary stars.  The catalogue contains measurements for 64,250 pairs.}

\noindent{$\bf{LSPM}$: A catalogue of 61,977 stars north of the celestial equator with proper motions
greater than $0.15\arcsec$ per year as presented by \citet{LepineShara2005}.}

\noindent{Luyten devised several numbering systems for the white dwarf and high proper motion stars 
that he discovered:}

\begin{itemize}

\item{$\bf{LFT}$: A catalogue of stars whose proper motions were greater than $0.5\arcsec$ annually 
\citep{Luyten1955}.}

\item{$\bf{LHS}$: A summary catalogue of stars whose proper motions were greater than $0.5\arcsec$ 
annually \citep{Luyten1976, Luyten1979a}.}

\item{$\bf{LTT}$: A catalogue of stars in the southern hemisphere whose proper motions exceed 
$0.2\arcsec$ annually \citep{Luyten1957}.}

\item{$\bf{NLTT}$: The New Luyten Two Tenths catalogue \citep{Luyten1979b, Luyten1979c, Luyten1980a, 
Luyten1980b}; \citep{LuytenHughes1980} contains stars whose proper motions are greater than 
$0.2\arcsec$ annually.}

\end{itemize}

\noindent{$\bf{MARK}$: The naming situation for the star named Mark A and other stars in its sequence 
should have been explained long ago.  \citet{MarkarianLipovetskij1973} published a list of galaxies 
with ultraviolet continua; the object designated Markarian 509 was one of these galaxies.  
\citet{MagnitskayaSaakyan1976} studied the light variation of Markarian 509.  W. Wisniewski, knowing 
that this author was searching for blue stars as possible candidate standard stars, told him of the 
existence of a blue star near Markarian 509, star A in Magnitskaya and Saakyan's Markarian 509 
sequence, and provided a chart of the blue star.  This author arbitrarily named the star Mark A, and 
subsequent nearby stars Mark A1, Mark A2, and so on.  The star Mark A is the only star in common 
between Magnitskaya and Saakyan's and the current photometric sequences.  The stars named Mark A, 
etc. in this paper now have been correlated with modern catalogues; please see Table \ref{tab:table5}.  
Parenthetically, there exists still another sequence for Markarian 509 \citep{Miller1981}; there are 
no Miller sequence stars in common with this paper.}

\noindent{$\bf{NSV}$: The NSV terminology began with the New Catalogue of Suspected Variable Stars 
\citep{KukarkinKholopov1982}.  One can now most easily access variable and suspected variable star 
information by entering the Sternberg Astronomical Institute's Web page\footnote{see 
http://www.sai.msu.su}, clicking on the "GCVS Research Group" (General Catalogue of Variable Stars), 
and then going to the appropriate catalogue.}

\noindent{$\bf{Oost}$: The Oost star numbers arise from a study of proper motion stars in 97 of the 
Selected Areas \citep{Oosterhoff1936}.}

\noindent{$\bf{PG}$: The Palomar-Green (PG) star numbering system arose from a study of 
ultraviolet-excess objects by \citet{Green1986}.}

\noindent{$\bf{PHL}$: The PHL star names follow from a search for faint blue stars 
\citep{HaroLuyten1962}.}
  
\noindent{$\bf{Ross}$: The Ross star numbers arise from a series of papers in the Astronomical 
Journal by F.E. Ross on high proper motion stars. The papers appeared in the time interval 1925-1939.  
See \citet{Luyten1963b} for a summary.}

\noindent{$\bf{Rubin}$: A search for faint blue stars in the galactic anti-center \citep{Rubin1974}.}

\noindent{$\bf{SA}$: The Kapteyn Selected Areas (SA) were defined in 1915 and were distributed around 
the sky in declination zones separated by 15 degrees as a basis for early galactic structure studies 
(reviewed by \citet{BlaauwElvius1965}). Coordinates and magnitudes for stars in these 206 Selected 
Areas were published by \citet{PickeringKapteyn1918}, \citet{Pickering1923, Pickering1924}.  Charts 
for the SA~stars have been published by \citet{BrunVehrenberg1965}.}

\noindent{$\bf{WD}$: The WD numbering system exists for white dwarf stars.  Excellent online sources 
of information for white dwarf stars include J. Holberg's Web site\footnote{see 
http://procyon.lpl.arizona.edu/WD/} and G. McCook and E. Sion's Web site\footnote{see 
http://www.astronomy.villanova.edu/WDcatalog/index.html}.  A print description of the latter is in 
\citet{McCookSion1999}.}

\noindent{$\bf{Wolf}$: Wolf star numbers are stars catalogued by M. Wolf in his studies of high proper 
motion stars. These papers appeared in the Astronomische Nachrichten in the time interval 1919-1931.  
See \citet{Luyten1963b} for a summary.}

Expanded comments, in the sense of increasing right ascension, follow for individual stars that 
appear in Table \ref{tab:table2}. Immediately following each star name is given, in brackets, an 
abbreviated right ascension and declination for that star, the idea being that one has an additional 
way of knowing the location on the sky while searching down the page for a star name.

\noindent{T~Phe~B [00:30:16; $-$46:27:59] = RW~Phe = AN~409.1929 = CD$-$47~128 = GSC~08024-00363.  
See note 1, Table 2 in \citet{Landolt1992}.  RW~Phe is a 5.4129 day eclipsing variable star 
discovered by \citet{Dartayet1929}.  T~Phe~B is sequence star ``h" for variable star T~Phe in 
\citet{FlemingPickering1907} and \citet{CampbellPickering1913}.}

\noindent{T~Phe~J [00:30:23; $-$46:23:52] = NLTT~290-35}

\noindent{T~Phe~F [00:30:50; $-$46:33:24] = GSC~08024-00830 = NSV~184.  See comment by 
\citet{Dartayet1929} that the T~Phe sequence star ``h1" might be variable.  The sequence is in 
\citet{FlemingPickering1907} and \citet{CampbellPickering1913}.  The AAVSO (d) chart for T~Phe 
(002546), plotted at a scale of $20\arcsec$ = 1 mm, is based on \citet{FlemingPickering1907} and 
\citet{CampbellPickering1913}. T~Phe~F here is the star marked as 132 on the AAVSO chart.}

\noindent{T~Phe~G [00:31:04; $-$46:22:51] = CD$-$47~134 = GSC~08024-00627}

\noindent{HD~2892 [00:32:12; $+$01:11:17] = BD$+$00~71 = GSC~00010-00965}

\noindent{BD$-$15~115 [00:38:20; $-$14:59:54] = GSC~05271-01721 = GCRV~20253}

\noindent{PG~0039+049 [00:42:05; $+$05:09:44] = GSC~00017-00283 = WD~0039$+$04}

\noindent{BD$-$11~162 [00:52:15; $-$10:39:56] = GD~743 = GSC~05270-01692 = NSV~15190.  
\citet{Zwicky1957} thought that the spectrum might be composite.  \citet{Klemola1962} finds 
$V=11.23$.  \citet{GreensteinEggen1966} indicated the spectrum to be composite, Op+G.  Also see 
\citet{GreensteinSargent1974}.  \citet{Dworetsky1982} found a spectral type of sdO+G and $V=11.19$.  
Since their color indices differed by $\sim0.05$ magnitudes from \citet{GreensteinSargent1974}, they 
suggested possible variability.  That suggestion may be supported by the comment of 
\citet{Klemola1962} that the spectrum looked like that of the Be star HD~14605.}

\noindent{SA~92-312 [00:53:16; $+$00:48:28] = BD$+$00~135 = GSC~00012-00120.  Note that the 
identification chart and photometry for this star is correct in \citet{Landolt1992}, but the star 
was mistakenly labeled as SA92-235, and the incorrect declination appears in Table 2 of 
\citet{Landolt1992}.}

\noindent{SA~92-339 [00:55:03; $+$00:44:11] = GSC~00012-00212 = NSV~15199 \citep{Landolt1992}}

\noindent{SA~92-188 [00:55:10; $+$00:23:12] = GSC~00012-00957 = NSV~15200 \citep{Landolt1992}}

\noindent{SA~92-409 [00:55:14; $+$00:56:07] = BD$+$00~143 = G~1-31 = G~70-20 = GSC~00012-00042 = 
LTT~10323 = NLTT~3047}

\noindent{Feige~11 [01:04:22; $+$04:13:37] = GSC~00022-00821 = PG~0101+039 = WD~0101$+$039}

\noindent{Feige~16 [01:54:08; $-$06:42:54] = GSC~04688-00087 = PHL~8054}

\noindent{G~3-33 [02:00:09; $+$13:04:04] = G~73-12 = GSC~00629-01255 = LFT~171 = LHS~11 = TZ~Ari, a 
flare star}

\noindent{Feige~22 [02:30:17; $+$05:15:51] = GSC~00052-00736 = WD~0227$+$050 = EGGR~19 = GCRV~53109 = 
NSV~843}

\noindent{PG~0231+051 [02:33:41; $+$05:18:40] = WD~0231$+$050}

\noindent{Feige~24 [02:35:08; $+$03:43:57] = GSC~00049-00886 = PG~0232+035 = WD~0232$+$035 = NSV~864 
= FS~Cet}

\noindent{BD$-$2~524 [02:57:40; $-$01:59:49] = Feige~29 = GCRV~53619 = GSC~04700-00656}

\noindent{GD~50 [03:48:50; $-$00:58:33] = EGGR~288 = WD~0346$-$011 = GSC~04717-00588}

\noindent{SA~95-301 [03:52:41; $+$00:31:21] = BD$+$00~669p = GSC~00066-01075}

\noindent{SA~95-302 [03:52:42: +00:31:18] = BD$+$00~669s = GSC~00066-01159}

\noindent{SA~95-96 [03:52:54; $+$00:00:19] = BD$-$00~613 = HD~24401 = GSC~00066-00944}

\noindent{SA~95-107 [03:53:25; $+$00:02:18] = FASTT~89}

\noindent{SA~95-43 [03:53:49; $-$00:03:01] = BD$-$00~615 = NSV~1400 = GSC~04718-00282}

\noindent{SA~95-137 [03:55:04; $+$00:03:33] = GSC~00066-00895 = USNO-B1.0~0900-0039856}

\noindent{SA~95-139 [03:55:05; $+$00:03:13] = GSC~00066-00931 = USNO-B1.0~0900-0039860}

\noindent{G~97-42 [05:28:01; $+$09:39:07] = Ci~20~329 = LHS~1761 = G~102-3 = LSPM~J0528+0938 = 
GSC~00704-00495 = LFT~411 = Ross~41 = GJ~203 = NLTT~15107}

\noindent{G~102-22 [05:42:05; $+$12:30:14] = LHS~31 = GSC~00722-00455 = V1352~Ori = LFT~425 = 
LTT~11704 = Ross~47 = Ci~20~344}

\noindent{GD~71 [05:52:28; $+$15:53:13] = EGGR~210 = LTT~11733 = WD~0549$+$158}

\noindent{SA~97-351 [05:57:37; $+$00:13:42] = BD$+$00~1231 = HD~290984 = GSC~00117-00727}

\noindent{SA~97-284 [05:58:25; $+$00:05:12] = BD$+$00~1237 = GSC~00117-01044}

\noindent{SA~98-978 [06:51:34; $-$00:11:28] = HD~292561 = GSC~04800-00469}

\noindent{SA~98-185 [06:52:02; $-$00:27:21] = BD$-$00~1466 = HD~292574 = GSC~04800-00923}

\noindent{SA~98-193 [06:52:04; $-$00:27:18] = HD~292575 = GSC~04800-01475}

\noindent{SA~98-653 [06:52:05; $-$00:18:19] = BD$-$00~1467 = HD~50188 = GSC~04800-01727}

\noindent{SA~99-6 [07:53:33; $-$00:49:37] = GSC~04833-00626 = NSV~3792}

\noindent{SA~99-408 [07:55:13; $-$00:25:33] = BD$-$00~1856 = GSC~04833-00597}

\noindent{SA~99-438 [07:55:54; $-$00:16:51] = BD$+$00~2129 = HD~64854 = GSC~04833-01301}

\noindent{SA~99-447 [07:56:07; $-$00:20:43] = BD$+$00~2131 = HD~64887 = GSC~04833-01280}

\noindent{SA~100-241 [08:52:35; $-$00:39:48] = BD$-$00~2081 = GSC~04865-00136}

\noindent{SA~100-162 [08:53:15; $-$00:43:29] = BD$-$00~2084 = GSC~04865-00508}

\noindent{PG~0918+029 [09:21:28; $+$02:46:03] = WD~0918$+$029 = GSC~00231-01572, a spectroscopic 
binary}

\noindent{BD$-$12~2918 [09:31:18; $-$13:29:20] = Ci~20~533 = Ross~440 = GCRV~6166 = GJ~352 = 
GSC~05472-00348 = IDS~09265$-$1303~AB = LHS~2151 = NLTT~21974 = NSV~4515}

\noindent{SA~101-324 [09:55:57; $-$00:23:14] = BD$+$00~2586 = GSC~04896-01224}

\noindent{SA~101-268 [09:56:19; $-$00:31:55] = G~161-88 = GSC~04896-01184 = LTT~18091 = Oost~455 = 
NLTT~22984, a high proper motion star}

\noindent{SA~101-363 [09:58:19; $-$00:25:35] = BD$+$00~2593 = HD~86408 = GSC~04896-01300}

\noindent{GD~108 [10:00:47; $-$07:33:31] = PG~0958-073 = WD~0958$-$073 = GSC~05476-00372}

\noindent{BD$+$1~2447 [10:28:56; $+$00:50:28] = Ci~20~580 = Ross~446 = GCRV~6576 = G~ 55-24 = 
G~162-60 = GJ~393 = GSC~00246-01068 = LFT~719 = NLTT~24467, a high proper motion star}

\noindent{G~162-66 [10:33:43; $-$11:40:39] = EGGR~70 = LTT~3870 = GSC~05495-00166 = 
NLTT~24689 = WD~1031$-$115, a white dwarf}

\noindent{G~44-27 [10:36:02; $+$05:07:11] = G~55-33 = GJ~398 = LHS~2285 = LSPM~J1036+0507 = 
GSC~00259-00166 = NLTT~24797 = RY~Sex, a flare star}

\noindent{PG~1034+001 [10:37:04; $-$00:08:20] = WD~1034$+$001 = GSC~04912-00085, a 
white dwarf}

\noindent{G~163-6 [10:42:55; $+$02:47:22] = NLTT~25129 = GSC~00257-00154 = LSPM~J1042+0247, a high 
proper motion star}

\noindent{PG~1047+003 [10:50:03; $-$00:00:32] = GSC~04914-00003 = UY~Sex, a variable star}

\noindent{G~44-40 [10:50:54; $+$06:48:57] = Ci~20~591 = GCRV~6733 = G~45-8 = LHS~294 = GJ~402 = 
Wolf~358 = GSC~00261-00224 = NLTT~25500 = J1050+0648 = EE~Leo, a high proper motion star}

\noindent{SA~102-620 [10:55:06; $-$00:48:19] = BD$-$00~2387 = Ross~898 = G~163-22 = GSC~04914-00700 = 
NLTT~25717, a high proper motion star}

\noindent{G~45-20 [10:56:38; $+$07:02:23] = Ci~20~600 = Wolf~359 = GCRV~6780 = GJ~406 = 
GSC~00261-00377 = NLTT~25782 = CN~Leo = LSPM~J1056+0700, a flare star and a high proper motion star}

\noindent{SA~102-1081 [10:57:04; $-$00:13:10] = BD$+$00~2717 = GSC~04914-00950}

\noindent{G~163-27 [10:57:35; $-$07:31:23] = EGGR~74 = GSC~05500-01023 = LHS~2333 = GJ~1140 = 
NLTT~25836 = PG~1055-073 = WD~1055$-$072, a white dwarf}

\noindent{G~163-50 [11:08:00; $-$05;09:26] = EGGR~76 = GSC~04927-00597 = WD~1105$-$048 = NLTT~26379 = 
PG~1105-049 = IDS~11030$-$0436~A = GJ~1142~B = NSV~5096; a white dwarf}

\noindent{G~163-51 [11:08:07; $-$05:13:47] = GJ~1142~A = GSC~04927-01272 = NLTT~26385 = 
IDS~11029$-$0436~B, a high proper motion star}

\noindent{BD +5 2468 [11:15:31; $+$04:57:24] = HD~97859 = GSC~00266-00468 = NSV~18709, however the 
mean error of a single observation in Table \ref{tab:table2} only is $0.011$ magnitude}

\noindent{HD~100340 [11:32:50; $+$05:16:36] = BD$+$06~2461 = GSC~00277-00509}

\noindent{BD$+$5~2529 [11:41:50; $+$05:08:27] = Ross~911 = LHS~6212 = GCRV~62701 = G~11-14 = G~10-43 
= GJ~9372 = NLTT~28274 = GSC~00278-00150, a high proper motion star}

\noindent{G~10-50 [11:47:44; $+$00:48:55] = Ci~20~662 = Ross~128 = GCRV~7139 = G~11-22 = GJ~447 = 
NLTT~28570 = GSC~00272-00665 = FI~Vir, a flare star and a high proper motion star}

\noindent{SA~103-302 [11:56:06; $-$00:47:54] = BD$+$00~2860 = HD~103646 = GSC~04932-00143}

\noindent{SA~103-526 [11:56:54; $-$00:30:13] = BD$+$00~2862 = GSC~04932-00246}

\noindent{G~12-43 [12:33:20; $+$09:01:08] = Ci~20~716 = Wolf~424 = G~60-14 = GJ~473 = GSC~00874-00306 
= NLTT~31083 = LHS~333 = GCRV~7553 = LSPM~J1233+0901 = FL Vir, a flare star and a high proper motion 
star}

\noindent{SA~104-306 [12:41:04; $-$00:37:11] = BD$+$00~2969 = HD~110281 = GSC~04949-00729 = KR~Vir, a 
semi-regular pulsating star}

\noindent{SA~104-461 [12:43:07; $-$00:32:21] = BD$+$00~2975 = HD~110572 = GSC~04949-01047 = NSV~5889}

\noindent{PG~1323-086 [13:25:39; $-$08:49:16] = WD~1323$-$085 = GSC~05544-00284}

\noindent{PG~1323-086A [13:25:50; $-$08:50:24] = GSC~05544-00493 = NSV~19792 discovered to be 
variable by \citet{Landolt1992} and independently by \citet{Schmidtke1990}}

\noindent{G~14-55 [13:28:22; $-$02:21:28] = Ci~20~780 = Ross~486A = G~62-42 = GSC~04959-00178 = 
GCRV~7989 = IDS~13231$-$0151~A = NLTT~34200 = G~14-55A = GJ~512~A = NSV~6261}

\noindent{SA~105-505 [13:35:25; $-$00:23:38] = BD$+$00~3077 = GCRV~64221 = GJ~1173 = GJ~9413 = 
GSC~04966-01321 = NLTT~34546 = NSV~6329}

\noindent{SA~105-815 [13:40:04; $-$00:02:10] = G~64-12 = GSC~04967-00579 = NLTT~34822 = Wolf~1492}

\noindent{BD$+$2~2711 [13:42:19; $+$01:30:19] = GSC~00308-00229}

\noindent{UCAC2~32376437 [13:42:23; $+$01:30:26] This relatively bright star is just a bit too faint 
to appear in any of the older catalogues.}

\noindent{HD~121968 [13:58:52; $-$02:55:12] = BD$-$02~3766 = GSC~04971-00048}

\noindent{PG~1407-013 [14:10:26; $-$01:30:17] = GSC~04976-00391}

\noindent{SA~106-1024 [14:40:07; $+$00:01:45] = GSC~00326-01010 = IP~Vir, a Delta Scuti star 
discovered by \citet{Landolt1990}}

\noindent{SA~106-700 [14:40:52; $-$00:23:36] = BD$+$00~3222 = GSC~04985-00500}

\noindent{SA~106-575 [14:41:38; $-$00:26:02] = BD$+$00~3224 = GSC~04985-00517}

\noindent{SA~106-485 [14:44:14; $-$00:37:07] = BD$+$00~3229 = HD~129727 = GSC~04985-00159}

\noindent{PG~1514+034 [15:17:14; $+$03:10:27] = EGGR~440 = GJ~3899 = WD~1514$+$033 = GSC~00340-00678}

\noindent{PG~1525-071 [15:28:11; $-$07:16:27] = GSC~05015-00248}

\noindent{PG~1528+062 [15:30:50; $+$06:00:56] = GSC~00362-00102}

\noindent{PG~1530+057 [15:33:11; $+$05:32:27] = GSC~00358-00731}

\noindent{SA~107-544 [15:36:48; $-$00:15:07] = BD$+$00~3379 = HD~139197 = GSC~05016-00561}

\noindent{SA~107-347 [15:38:36; $-$00:35:58] = BD$-$00~2991 = HD~139513 = GSC~05017-00367}

\noindent{G~153-41 [16:17:55; $-$15:35:52] = EGGR~118 = NLTT~42430 = WD~1615$-$157 = GSC~06202-00265 
= LTT~6497 = Oost~596}

\noindent{G~138-25 [16:25:14; $+$15:41:15] = LHS~418 = GSC~01506-01012 = NLTT~42743, a high proper 
motion star}

\noindent{BD$-$12~4523 [16:30:18; $-$12:39:08] = Ci~20~995 = GCRV~9490 = G~153-58 = GJ~628 = 
GSC~05635-00564 = LHS~419 = NLTT~42927 = Wolf~1061 = NSV~7768 = V2306~Oph, a high proper motion star}

\noindent{HD~149382 [16:34:23; $-$04:00:52] = BD$-$03~3967 = GCRV~9530 = GSC~05056-00274 = 
PG~1631-039}

\noindent{SA~108-1332 [16:35:21; $-$00:04:05] = BD$+$00~3552 = HD~149506 = GSC~05048-00853 = 
IDS~16302$+$0008~A}

\noindent{PG~1633+099 [16:35:24; $+$09:47:50] = GSC~00964-01526 = BPS~CS~22878-0031}

\noindent{SA~108-1491 [16:37:14; $-$00:02:42] = BD$+$00~3554 = HD~149825 = GSC~05048-00933}

\noindent{SA~108-551 [16:37:47; $-$00:33:06] = BD$-$00~3152 = GSC~05049-00473}

\noindent{Wolf~629 [16:55:27; $-$08:18:53] = BD$-$08~4352C = GCRV~9744 = GJ~643 = LHS~427 = 
GSC~05642-01473 = IDS~16501$-$0809~C = NLTT~43797, a high proper motion star}

\noindent{PG~1657+078 [16:59:32; $+$07:43:31] = GSC~00976-01448}

\noindent{BD$-$4~4226 [17:05:15; $-$05:05:05] = HD~154363B = Ci~20~1018 = Wolf~636 = GCRV~9854 = 
G~19-14 = GJ~654 = GSC~05072-00347 = IDS~16597$-$0456~B = NLTT~44131 = LHS~432 = NSV~8176, a high 
proper motion star}

\noindent{SA~109-231 [17:45:20; $-$00:25:51] = BD$-$00~3353 = GSC~05082-02064}

\noindent{SA~109-537 [17:45:42; $-$00:21:34] = BD$-$00~3356 = GSC~05082-01171}

\noindent{G~21-15 [18:27:13; $+$04:03:05] = EGGR~125 = GCRV~68202 = G~141-11 = NLTT~46505 = Ross~137
= WD~1824$+$040 = LSPM~J1827+0403, = LSPM~J1827+0403, a white dwarf}

\noindent{SA~110-340 [18:41:29; $+$00:15:22] = BD$+$00~3992 = HD~172652 = GSC~00447-00541}

\noindent{SA~110-246 [18:41:51; $+$00:05:20] = GSC~00447-00424 = NLTT~46871, a high proper motion 
star}

\noindent{SA~111-773 [19:37:16; $+$00:10:59] = BD$-$00~3800 = HD~185025 = GSC~00478-01575}

\noindent{SA~111-1969 [19:37:44; $+$00:25:48] = BD$+$00~4260 = GSC~00479-01376}

\noindent{SA~112-275 [20:42:36; $+$00:07:21] = BD$-$00~4073 = GSC~00511-02062}

\noindent{Wolf~918 [21:11:30; $-$13:08:22] = Ci~20~1261 = GJ~821 = NLTT~50633 = GSC~05783-00134 = 
LHS~65, a high proper motion star}

\noindent{G~26-7 [21:31:19; $-$09:47:26] = Ci~20~1288 = Wolf~922 = GCRV~72228 = GJ~831 = 
GSC~05790-00182 = NLTT~51428 = NSV~13753 = BB~Cap}

\noindent{SA~113-466 [21:41:28; $+$00:40:14] = BD$+$00~4766 = GSC~00543-00227}

\noindent{SA~113-475 [21:41:51; $+$00:39:19] = BD$+$00~4767 = GSC~00543-00262}

\noindent{SA~113-492 [21:42:28; $+$00:38:21] = GSC~00543-01655 = IDS~21373$+$0010~B}

\noindent{SA~113-493 [21:42:29; $+$00:38:10] = GSC~00543-01668 = IDS~21373$+$0010~A}

\noindent{SA~113-495 [21:42:30; $+$00:38:07] = GSC~00543-00542 = IDS~21373$+$0010~C}

\noindent{G~93-48 [21:52:25; $+$02:23:20] = EGGR~150 = GJ~838.4 = NLTT~52306 = WD~2149$+$021 = 
GSC~00548-00105 = LSPM~J2152+0223, a white dwarf}

\noindent{PG~2213-006 [22:16:28; $-$00:21:15] = GSC~05225-00812}

\noindent{G~156-31 [22:38:28; $-$15:19:17] = GCRV~14217 = GJ~866 = GJ~866~A = LHS~68 = 
GSC~06386-00505 = NLTT~54407 = EZ~Aqr, a flare star}

\noindent{SA~114-755 [22:42:08; $+$01:16:49] = BD$+$00~4910 = GSC~00568-00923}

\noindent{SA~114-176 [22:43:11; $+$00:21:16] = BD$-$00~4408 = HD~215141 = GSC~00568-01711}

\noindent{HD~216135 [22:50:28; $-$13:18:44] = BD$-$14~6357 = Feige~107 = GSC~05819-00618}

\noindent{G~156-57 [22:53:16; $-$14:15:38] = BD$-$15~6290 = Ci~20~1387 = GCRV~14365 = G~156-57A = 
GJ~876 = GJ~876~A = GSC~05819-00957 = LHS~530 = NLTT~55130 = LHS~530, a high proper motion star; = 
IL~Aqr}

\noindent{GD~246 [23:12:22; $+$10:47:04] = EGGR~233 = BPM~97895 = PG~2309+105 = WD~2309$+$105 = 
GSC~01164-01078, a white dwarf}

\noindent{Feige~108 [23:16:12; $-$01:50:35] = EGGR~157 = PG~2313-021 = GCRV~73935 = GSC~05243-00817 = 
WD~2313$-$021 = NSV~26050, a white dwarf}

\noindent{SA~115-271 [23:42:41; $+$00:45:10] = BD$-$00~4557 = GSC~00586-00979}

\noindent{SA~115-516 [23:44:15; $+$01:14:13] = BD$+$00~5040 = GSC~00586-00132}

\noindent{BD$+$1~4774 [23:49:13; $+$02:24:04] = Ci~18~3124 = GCRV~14913 = G~29-68 = G~31-6 = GJ~908 = 
GSC~00586-00610 = NLTT~58069 = NSV~14719 = BR~Psc, a flare star, and a high proper motion star}

\noindent{PG~2349+002 [23:51:53; $+$00:28:17] = GSC~00587-00489}

\acknowledgements{It always is a pleasure to acknowledge the staff of the CTIO for their hospitality 
and assistance!  Individuals always available to help in anyway include A. Alvarez, E. Cosgrove, M. 
Fernandez, A. Gomez, A. Guerra, R. Leiton, D. Maturana, J. Perez, S. Pizarro, M. Rodriguez, D. Rojas, 
O. Saa, N. Saavedra, E. Schmidt, H. Tirado, P. Ugarte, R. Venegas, and A. Zuniga.  I thank Brian 
McLean for making access possible for larger sized images.  Thanks go to Phil Massey who read a draft 
of this paper.  Brian Skiff updated the author with techniques to ensure that the coordinates and 
proper motions are modern and accurate.  The appearance of this paper's figures and tables are due to 
the skills of James L. Clem.  This observational program has been and is supported by NSF grants 
AST-9528177, AST-0097895, AST-0503871, and AST-0803158.}



\newpage
\clearpage
\input{table01.tex}

\clearpage

\newpage
\clearpage
\input{table02.tex}
\clearpage

\newpage
\clearpage
\input{table03.tex}

\clearpage

\newpage
\clearpage
\input{table04.tex}
\clearpage

\newpage
\clearpage
\input{table05.tex}

\clearpage


\newpage
\clearpage
\input{figures.tex}

\clearpage

\end{document}

%% file: table01.tex
\begin{deluxetable}{lrccccr}
\tablecaption{Non-linear coefficients.}
\tabletypesize{\scriptsize}
\tablewidth{0pt}
\tablehead{\colhead{Instrument}            &
           \colhead{Index}                 &
           \colhead{zero}                  &
           \colhead{error zero}            &
           \colhead{slope}                 &
           \colhead{error slope}           &
           \colhead{Color Range}           }
\startdata
T60RIPOS              & $(B-V)_{c}$ & $-$0.00212 & $\pm$0.00284 & $+$1.03758 & $\pm$0.01959 &      $(B-V)<+0.1$ \\
~~~(4 nights)         & $(B-V)_{c}$ & $+$0.00247 & $\pm$0.00288 & $+$1.00401 & $\pm$0.00537 & $+0.1<(B-V)<+1.0$ \\
                      & $(B-V)_{c}$ & $+$0.02200 & $\pm$0.00684 & $+$0.98357 & $\pm$0.00511 &      $(B-V)>+1.0$ \\ [1mm] \\
                      & $(U-B)_{c}$ & $-$0.07830 & $\pm$0.07168 & $+$0.88102 & $\pm$0.08268 &      $(U-B)<-0.4$ \\
                      & $(U-B)_{c}$ & $-$0.01327 & $\pm$0.00241 & $+$0.97997 & $\pm$0.01308 & $-0.4<(U-B)<+0.5$ \\
                      & $(U-B)_{c}$ & $-$0.00187 & $\pm$0.02028 & $+$1.00803 & $\pm$0.01427 &      $(U-B)>+0.5$ \\ [1mm] \\
\enddata
\label{tab:table1}
\end{deluxetable}

%% file: table02.tex
\begin{deluxetable}{p{0.7in}llrcccccrrcccccc}
\rotate
\tablecaption{$UBVRI$ Photometry of Standard Stars.}
\tabletypesize{\tiny}
\tablewidth{0pt}
\tablehead{\multicolumn{11}{c}{ }                            &
           \multicolumn{6}{c}{\sc{Mean Error of the Mean}}   \\ [1pt] \\ \cline{12-17} \\ [1pt]
           \colhead{\sc{Star}}                               &
           \colhead{$\alpha$ (J2000.0)}                      &
           \colhead{$\delta$ (J2000.0)}                      &
           \colhead{$V$}                                     &
           \colhead{$B-V$}                                   &
           \colhead{$U-B$}                                   &
           \colhead{$V-R$}                                   &
           \colhead{$R-I$}                                   &
           \colhead{$V-I$}                                   &
           \colhead{$n$}                                     &
           \colhead{$m$}                                     &
           \colhead{$V$}                                     &
           \colhead{$B-V$}                                   &
           \colhead{$U-B$}                                   &
           \colhead{$V-R$}                                   &
           \colhead{$R-I$}                                   &
           \colhead{$V-I$}                                   \\
           \colhead{(1)}                                     &
           \colhead{(2)}                                     &
           \colhead{(3)}                                     &
           \colhead{(4)}                                     &
           \colhead{(5)}                                     &
           \colhead{(6)}                                     &
           \colhead{(7)}                                     &
           \colhead{(8)}                                     &
           \colhead{(9)}                                     &
           \colhead{(10)}                                    &
           \colhead{(11)}                                    &
           \colhead{(12)}                                    &
           \colhead{(13)}                                    &
           \colhead{(14)}                                    &
           \colhead{(15)}                                    &
           \colhead{(16)}                                    &
           \colhead{(17)}                                    }
\startdata
TPhe I      &  00 30 04.593  &  $-$46 28 10.17  & 14.820 & $+$0.764 & $+$0.338 & $+$0.422 & $+$0.395 & $+$0.817 &  25 &  13 &  0.0026 &  0.0032 &  0.0072 &  0.0036 &  0.0098 &  0.0110   \\ 
TPhe A      &  00 30 09.594  &  $-$46 31 28.91  & 14.651 & $+$0.793 & $+$0.380 & $+$0.435 & $+$0.405 & $+$0.841 &  29 &  12 &  0.0028 &  0.0046 &  0.0071 &  0.0019 &  0.0035 &  0.0032   \\ 
TPhe H      &  00 30 09.683  &  $-$46 27 24.30  & 14.942 & $+$0.740 & $+$0.225 & $+$0.425 & $+$0.425 & $+$0.851 &  23 &  12 &  0.0029 &  0.0029 &  0.0071 &  0.0035 &  0.0077 &  0.0098   \\ 
TPhe B      &  00 30 16.313  &  $-$46 27 58.57  & 12.334 & $+$0.405 & $+$0.156 & $+$0.262 & $+$0.271 & $+$0.535 &  29 &  17 &  0.0115 &  0.0026 &  0.0039 &  0.0020 &  0.0019 &  0.0035   \\ 
TPhe C      &  00 30 16.98   &  $-$46 32 21.4   & 14.376 & $-$0.298 & $-$1.217 & $-$0.148 & $-$0.211 & $-$0.360 &  39 &  23 &  0.0022 &  0.0024 &  0.0043 &  0.0038 &  0.0133 &  0.0149   \\ 
TPhe D      &  00 30 18.342  &  $-$46 31 19.85  & 13.118 & $+$1.551 & $+$1.871 & $+$0.849 & $+$0.810 & $+$1.663 &  37 &  23 &  0.0033 &  0.0030 &  0.0118 &  0.0015 &  0.0023 &  0.0030   \\ 
TPhe E      &  00 30 19.768  &  $-$46 24 35.60  & 11.631 & $+$0.443 & $-$0.103 & $+$0.276 & $+$0.283 & $+$0.564 &  38 &  10 &  0.0017 &  0.0013 &  0.0025 &  0.0007 &  0.0016 &  0.0020   \\ 
TPhe J      &  00 30 23.02   &  $-$46 23 51.6   & 13.434 & $+$1.465 & $+$1.229 & $+$0.980 & $+$1.063 & $+$2.043 &  28 &  15 &  0.0023 &  0.0043 &  0.0059 &  0.0011 &  0.0015 &  0.0011   \\
\enddata
\label{tab:table2}
\end{deluxetable}

%% file: table03.tex
\newpage
\begin{deluxetable}{lcc}
\tablecaption{Error Analysis for the Standard Stars in Table \ref{tab:table2}.}
\tablewidth{0pt}
\tablehead{\colhead{ }                                    &
           \colhead{Mean Errors of a Single Observation}  &
           \colhead{Mean Errors of the Mean}              }
\startdata
$V$       &  $0.0144\pm0.0011$  &  $0.0036\pm0.0004$ \\
$B-V$     &  $0.0191\pm0.0022$  &  $0.0051\pm0.0008$ \\
$U-B$     &  $0.0492\pm0.0085$  &  $0.0143\pm0.0036$ \\
$V-R$     &  $0.0115\pm0.0008$  &  $0.0029\pm0.0003$ \\
$R-I$     &  $0.0166\pm0.0013$  &  $0.0040\pm0.0003$ \\
$V-I$     &  $0.0207\pm0.0015$  &  $0.0050\pm0.0004$ \\
\enddata
\label{tab:table3}
\end{deluxetable}

%% file: table04.tex
\begin{deluxetable}{lccccrrrl}
\rotate
\tablecaption{Field Centers for Sequences.}
\tabletypesize{\tiny}
\tablewidth{0pt}
\tablehead{\colhead{Field Name}                 &
           \colhead{$\alpha$ (J2000.0)}         &
           \colhead{$\delta$ (J2000.0)}         &
           \colhead{FOV}                        &
           \colhead{No. Stars}                  &
           \colhead{Range in $V$}               &
           \colhead{Range in $B-V$}             &
           \colhead{Range in $U-B$}             &
           \colhead{Comments}                   }
\startdata
TPhe         & 00:30:34.5 & $-$46:28:08 & $15x15$ & 11 & $10.45 ~\leftrightarrow ~ 14.94$ & $-0.30 ~\leftrightarrow ~ +1.55$ & $-1.22 ~\leftrightarrow ~ +1.91$ & all stars except TPhe K \\
SA92         & 00:55:15.5 & $+$00:47:26 & $70x70$ & 41 & $10.60 ~\leftrightarrow ~ 15.92$ & $+0.32 ~\leftrightarrow ~ +1.64$ & $-0.18 ~\leftrightarrow ~ +1.99$ & all stars except those with fewer than 5 measures \\
~~SA92 SF1   & 00:55:24.6 & $+$00:39:46 & $10x10$ &  4 & $11.62 ~\leftrightarrow ~ 15.07$ & $+0.43 ~\leftrightarrow ~ +1.16$ & $-0.04 ~\leftrightarrow ~ +1.11$ & SA92 342, $\it{259}$, 260, 263 \\
~~SA92 SF2   & 00:56:06.7 & $+$00:52:03 & $10x10$ &  4 & $13.94 ~\leftrightarrow ~ 14.96$ & $+0.57 ~\leftrightarrow ~ +1.19$ & $-0.04 ~\leftrightarrow ~ +1.20$ & SA92 425, 426, 355, 430 \\
~~SA92 SF3   & 00:57:01.9 & $+$00:40:21 & $10x10$ &  3 & $11.63 ~\leftrightarrow ~ 12.97$ & $+0.32 ~\leftrightarrow ~ +0.86$ & $-0.04 ~\leftrightarrow ~ +0.47$ & SA92 282, 364, 288 \\
F 11         & 01:04:25.0 & $+$04:12:31 & $10x10$ &  3 & $12.06 ~\leftrightarrow ~ 14.47$ & $-0.24 ~\leftrightarrow ~ +0.84$ & $-0.99 ~\leftrightarrow ~ +0.45$ & all stars \\
SA93         & 01:55:01.8 & $+$00:49:52 & $60x60$ &  4 & $11.55 ~\leftrightarrow ~ 12.01$ & $+0.49 ~\leftrightarrow ~ +1.08$ & $-0.05 ~\leftrightarrow ~ +0.93$ & all stars \\
PG0220+132   & 02:23:36.9 & $+$13:27:47 & $10x10$ &  3 & $14.22 ~\leftrightarrow ~ 15.77$ & $-0.13 ~\leftrightarrow ~ +0.94$ & $-0.92 ~\leftrightarrow ~ +0.32$ & with caution \\
\enddata
\label{tab:table4}
\end{deluxetable}

%% file: table05.tex
\begin{deluxetable}{p{0.75in}llccrrccc}
\tablecaption{Accurate Coordinates and Proper Motions for the Stars in Table \ref{tab:table2}.}
\tabletypesize{\tiny}
\tablewidth{0pt}
\tablehead{\multicolumn{5}{c}{ }                    &
           \colhead{$\mu_{\alpha}$}                 &
           \colhead{$\mu_{\delta}$}                 &
           \colhead{  }                             &
           \colhead{  }                             \\
           \colhead{Star Name}                      &
           \colhead{$\alpha$ (J2000.0)}             &
           \colhead{$\delta$ (J2000.0)}             &
           \colhead{UCAC2}                          &
           \colhead{2MASS-PSC}                      &
           \colhead{(mas yr$^{-1}$)}                &
           \colhead{(mas yr$^{-1}$)}                &
           \colhead{$\mu$ Ref.}                     }
\tablenotetext{~}{1.  \citep{Zacharias2004}[UCAC2]}
\tablenotetext{~}{2.  \citep{Monet2003}[USNO-B1.0 catalog]}
\tablenotetext{~}{3.  \citep{Klemola1987}}
\tablenotetext{~}{4.  \citep{LepineShara2005}}
\tablenotetext{~}{5.  \citep{Urban2004}}
\startdata
            &                &                  &            &                      &                      &                      &            \\
TPhe I      &  00 30 04.593  &  $-$46 28 10.17  &  11909220  & J00300459$-$4628102  &                13.6  &              $-$8.5  &         1  \\ 
TPhe A      &  00 30 09.594  &  $-$46 31 28.91  &  11679110  & J00300959$-$4631289  &                 8.3  &              $-$2.0  &         1  \\ 
TPhe H      &  00 30 09.683  &  $-$46 27 24.30  &  11909222  & J00300968$-$4627243  &                11.2  &              $-$8.1  &         1  \\ 
TPhe B      &  00 30 16.313  &  $-$46 27 58.57  &  11909226  & J00301631$-$4627586  &                 1.0  &              $-$5.4  &         1  \\ 
TPhe C      &  00 30 16.98   &  $-$46 32 21.4   &       ---  & J00301697$-$4632214  &   $-$24\phantom{.0}  &   $-$22\phantom{.0}  &         2  \\ 
TPhe D      &  00 30 18.342  &  $-$46 31 19.85  &  11679116  & J00301834$-$4631198  &                 2.3  &                 1.3  &         1  \\ 
TPhe E      &  00 30 19.768  &  $-$46 24 35.60  &  11909227  &                 ---  &                51.4  &                 1.6  &         1  \\
\enddata
\label{tab:table5}
\end{deluxetable}

%% file: figures.tex
\clearpage
\begin{figure}
\plotone{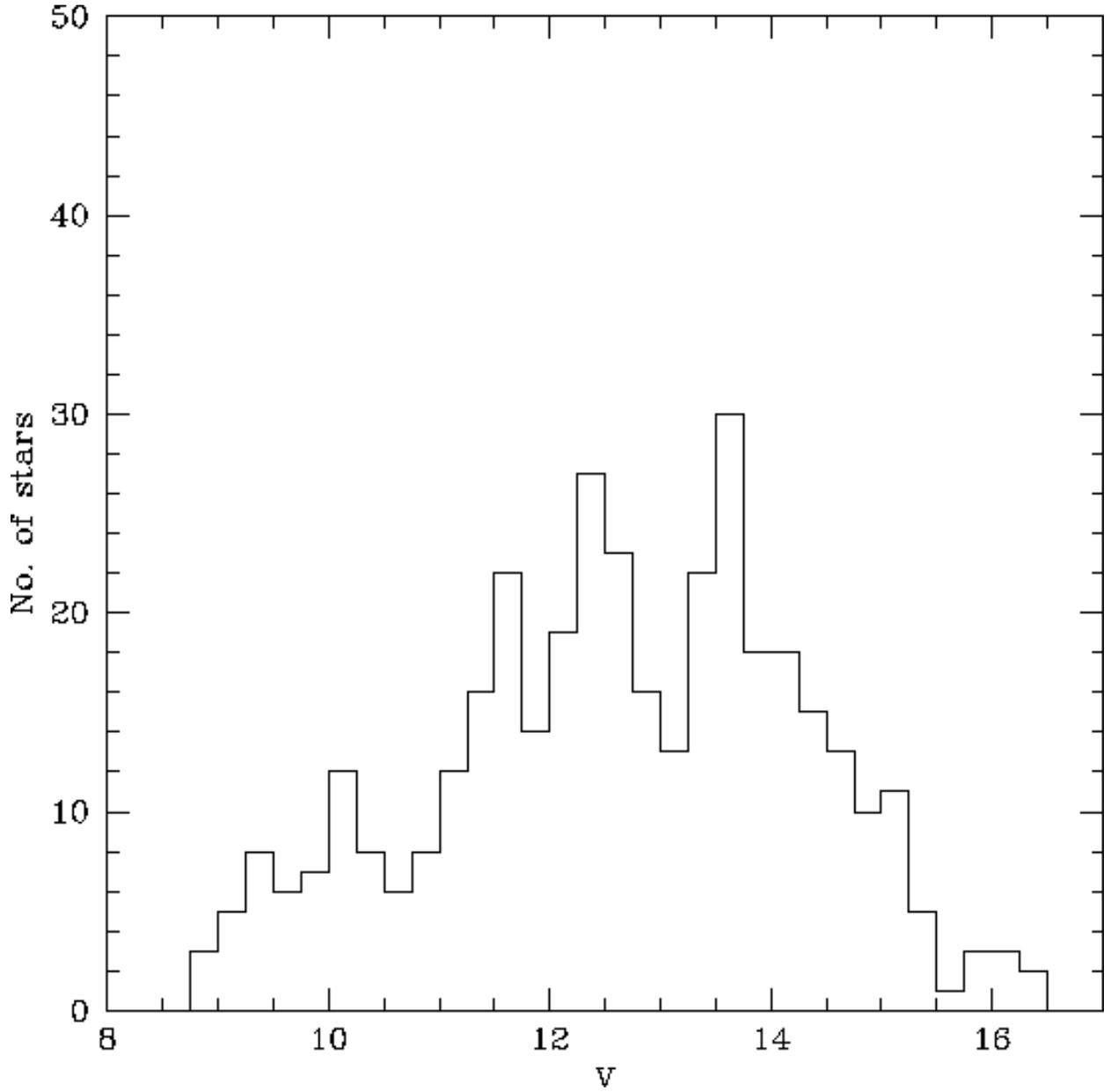}
\caption{Magnitude distribution for the standard stars listed in Table \ref{tab:table2} with five or more measures 
in intervals of 0.25 $V$ mag.}
\label{fig:figure137}
\end{figure}

\clearpage
\begin{figure}
\plotone{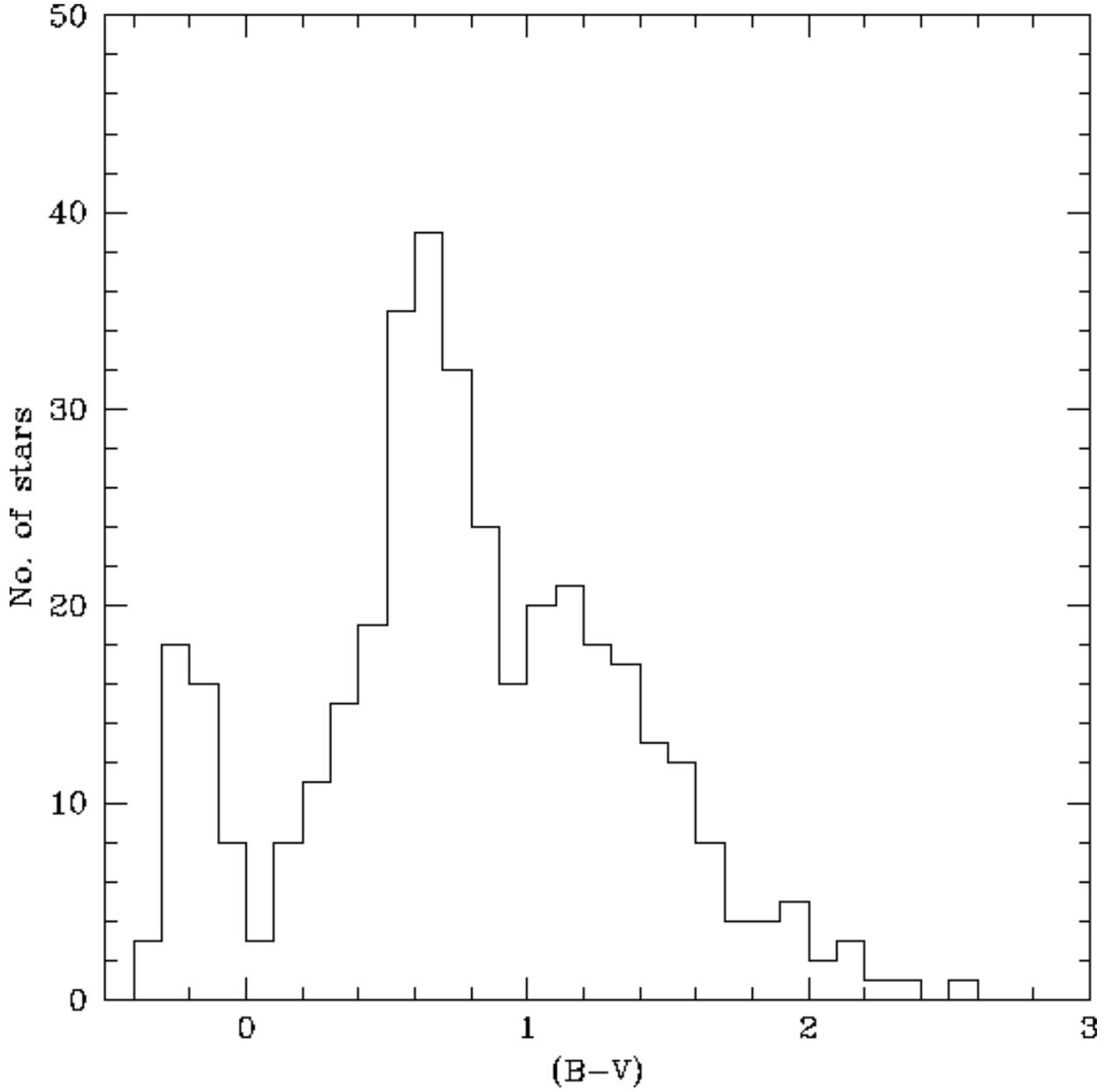}
\caption{Distribution in $(B-V)$ color index for the standard stars listed in Table \ref{tab:table2} with five or 
more measures in intervals of 0.1 mag.}
\label{fig:figure138}
\end{figure}

\clearpage
\begin{figure}
\plotone{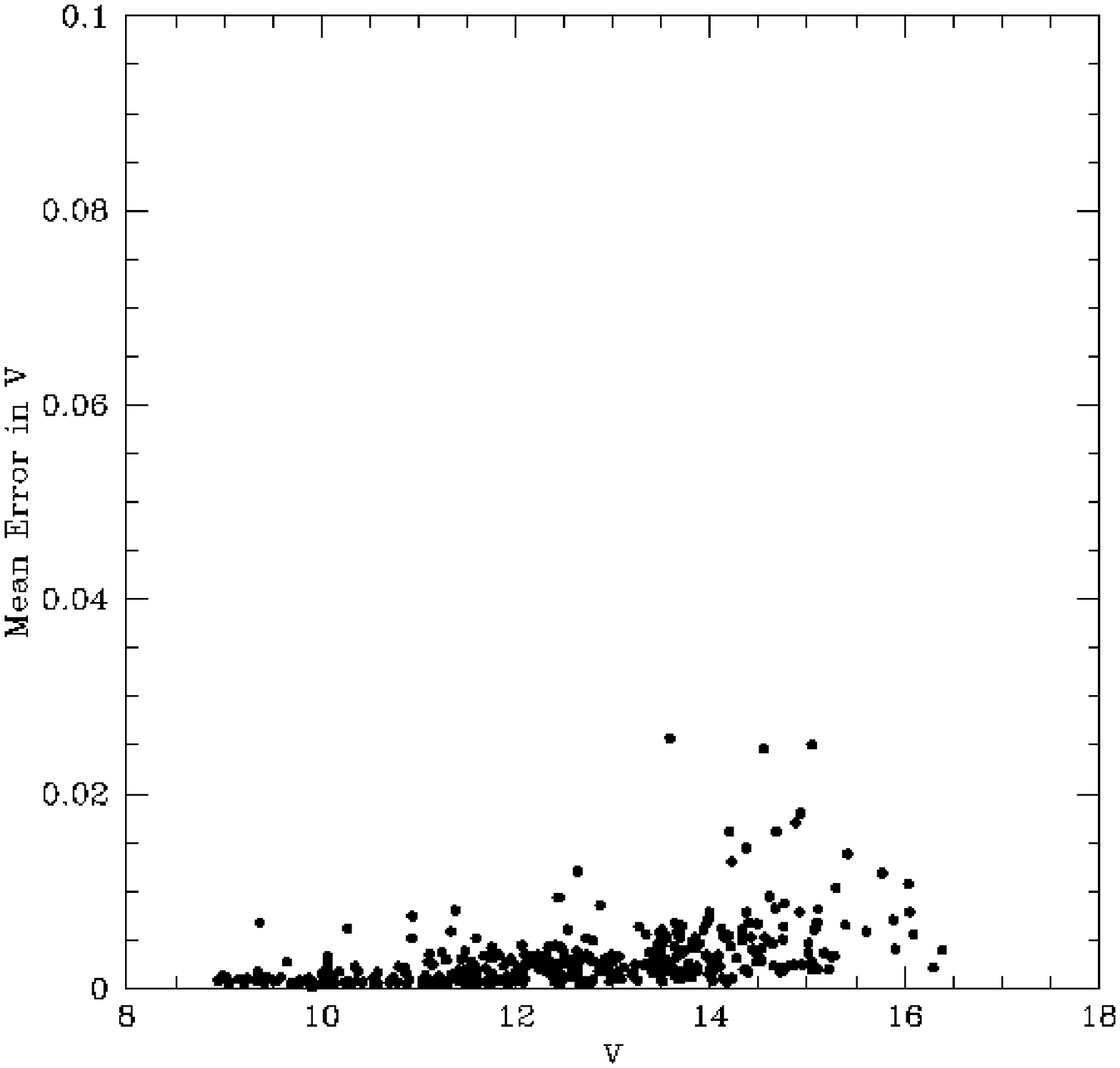}
\caption{Mean error of the mean of a single observation in $V$ for the standard stars as a function of $V$.}
\label{fig:figure139}
\end{figure}

\clearpage
\begin{figure}
\plotone{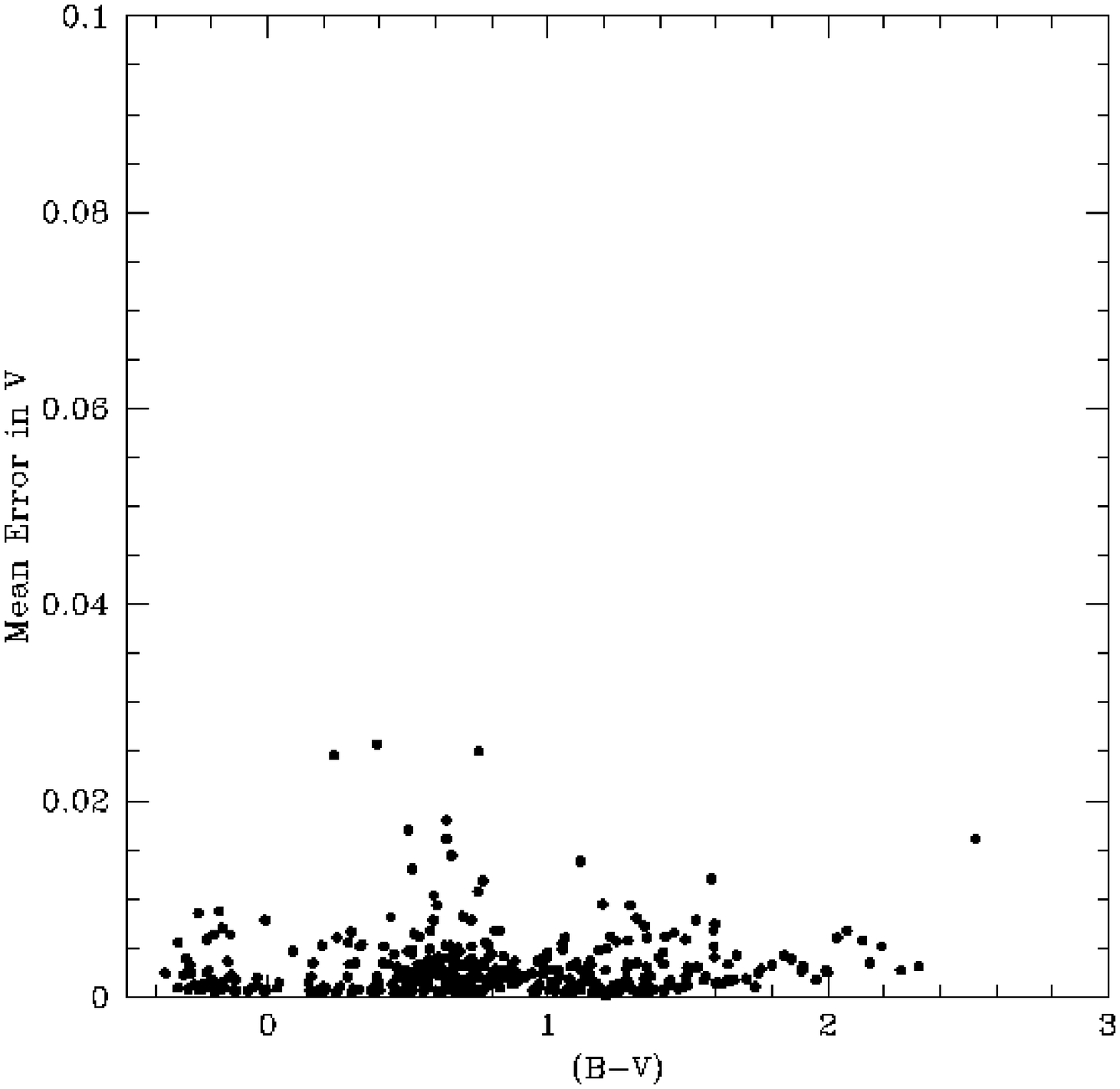}
\caption{Mean error of the mean of a single observation in $V$ for the standard stars as a function of $(B-V)$.}
\label{fig:figure140}
\end{figure}

\clearpage
\begin{figure}
\plotone{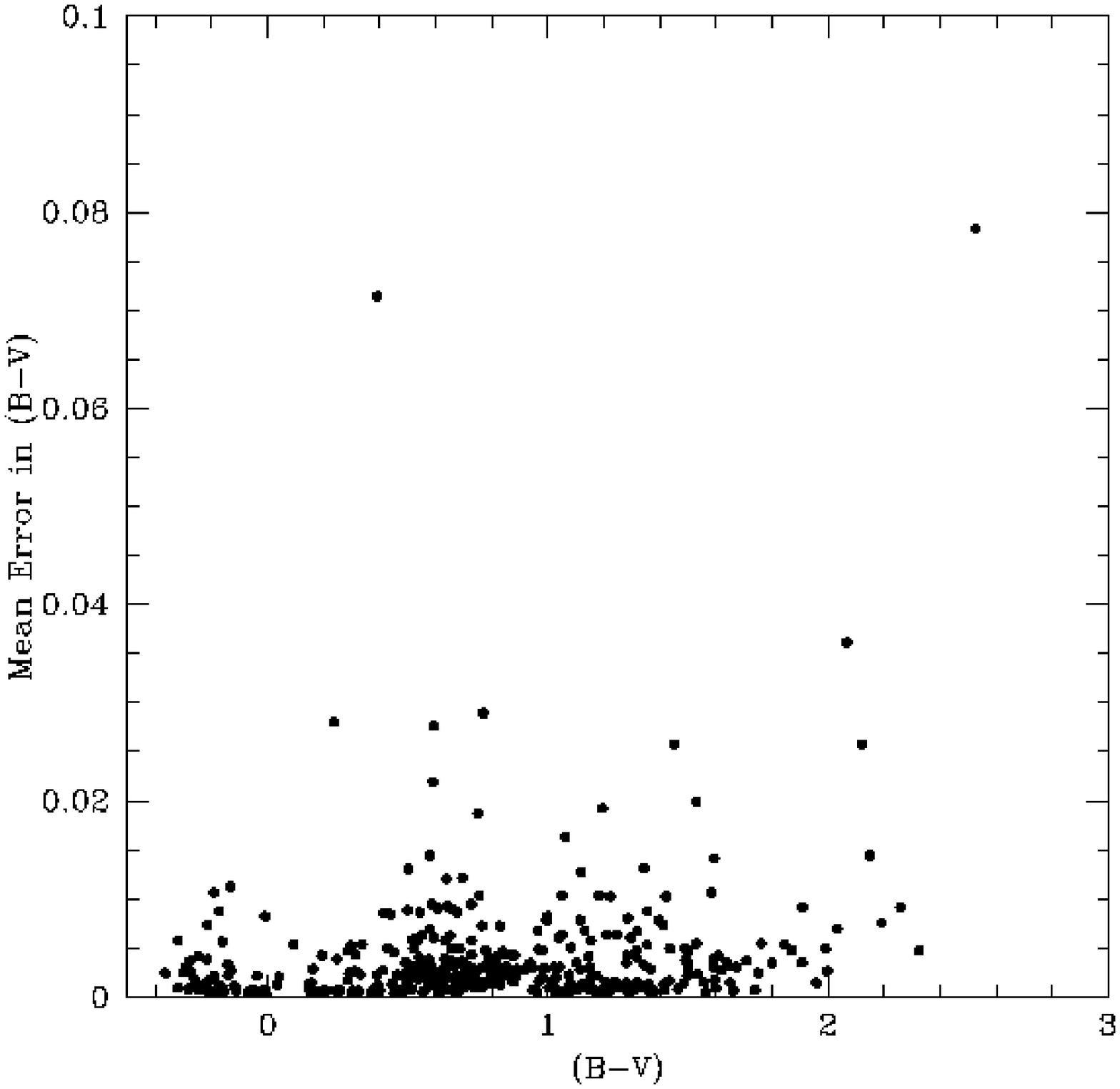}
\caption{Mean error of the mean of a single observation in $(B-V)$ for the standard stars as a function of $(B-V)$.}
\label{fig:figure141}
\end{figure}

\clearpage
\begin{figure}
\plotone{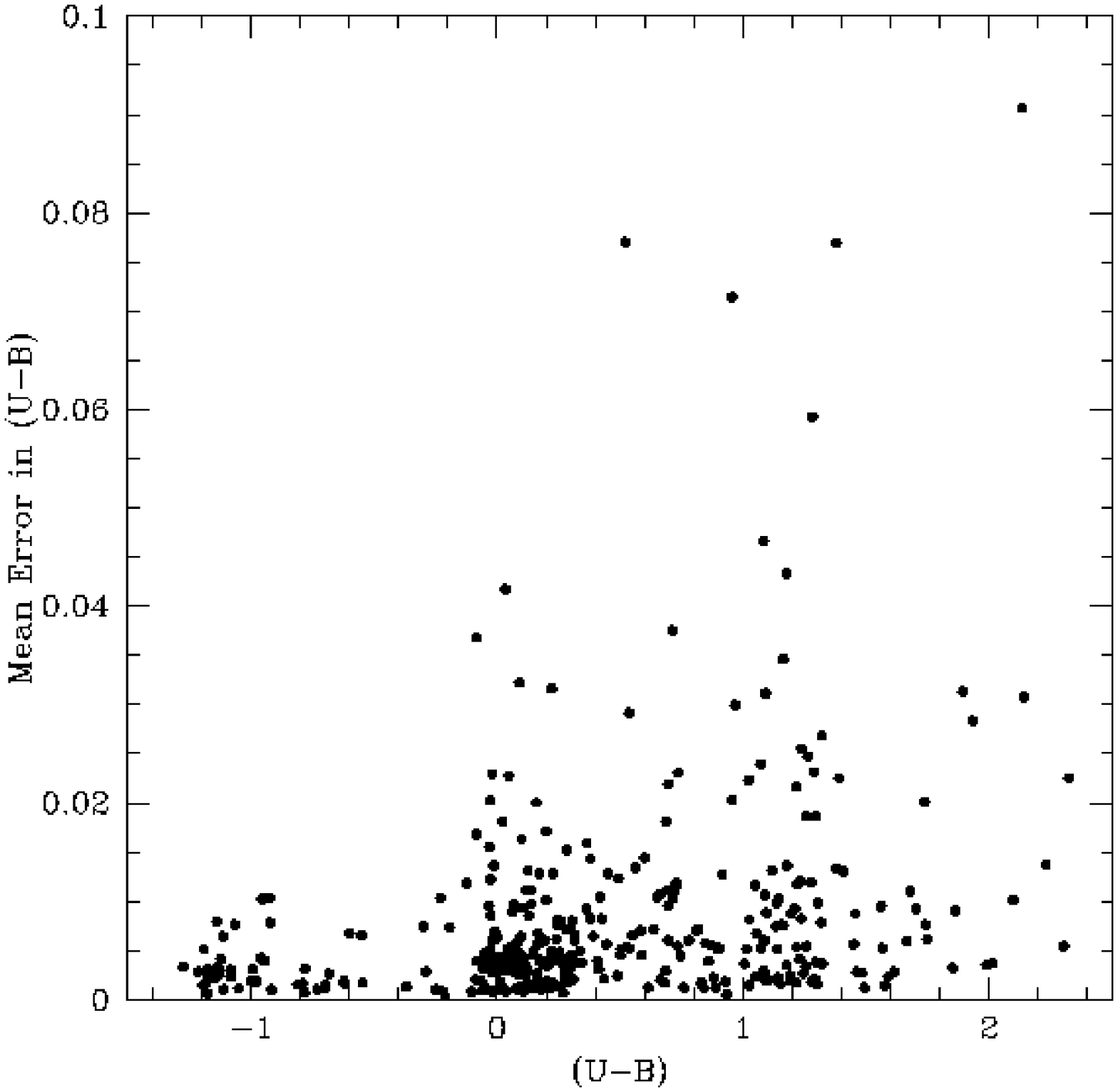}
\caption{Mean error of the mean of a single observation in $(U-B)$ for the standard stars as a function of $(U-B)$.}
\label{fig:figure142}
\end{figure}

\clearpage
\begin{figure}
\plotone{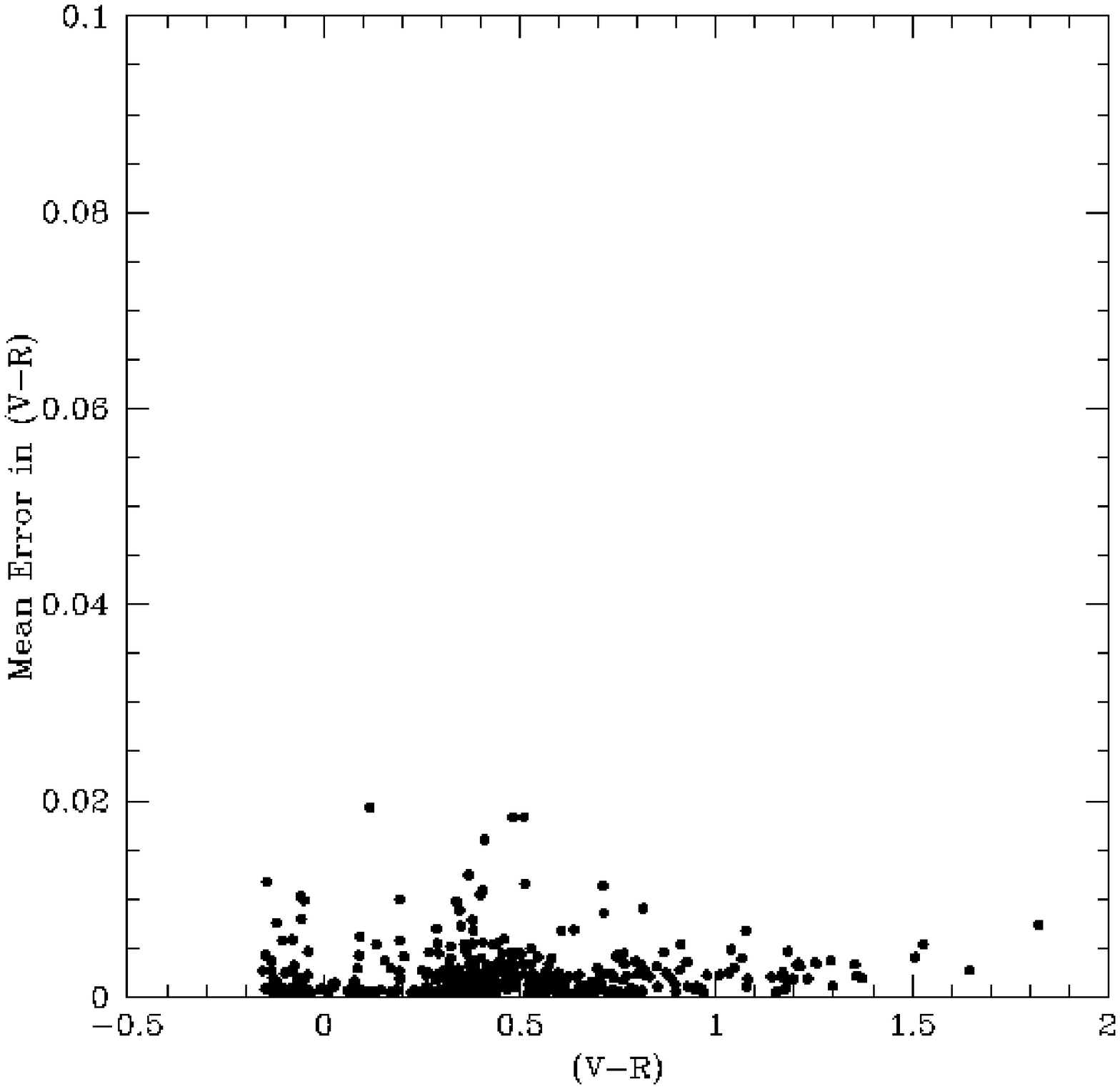}
\caption{Mean error of the mean of a single observation in $(V-R)$ for the standard stars as a function of $(V-R)$.}
\label{fig:figure143}
\end{figure}

\clearpage
\begin{figure}
\plotone{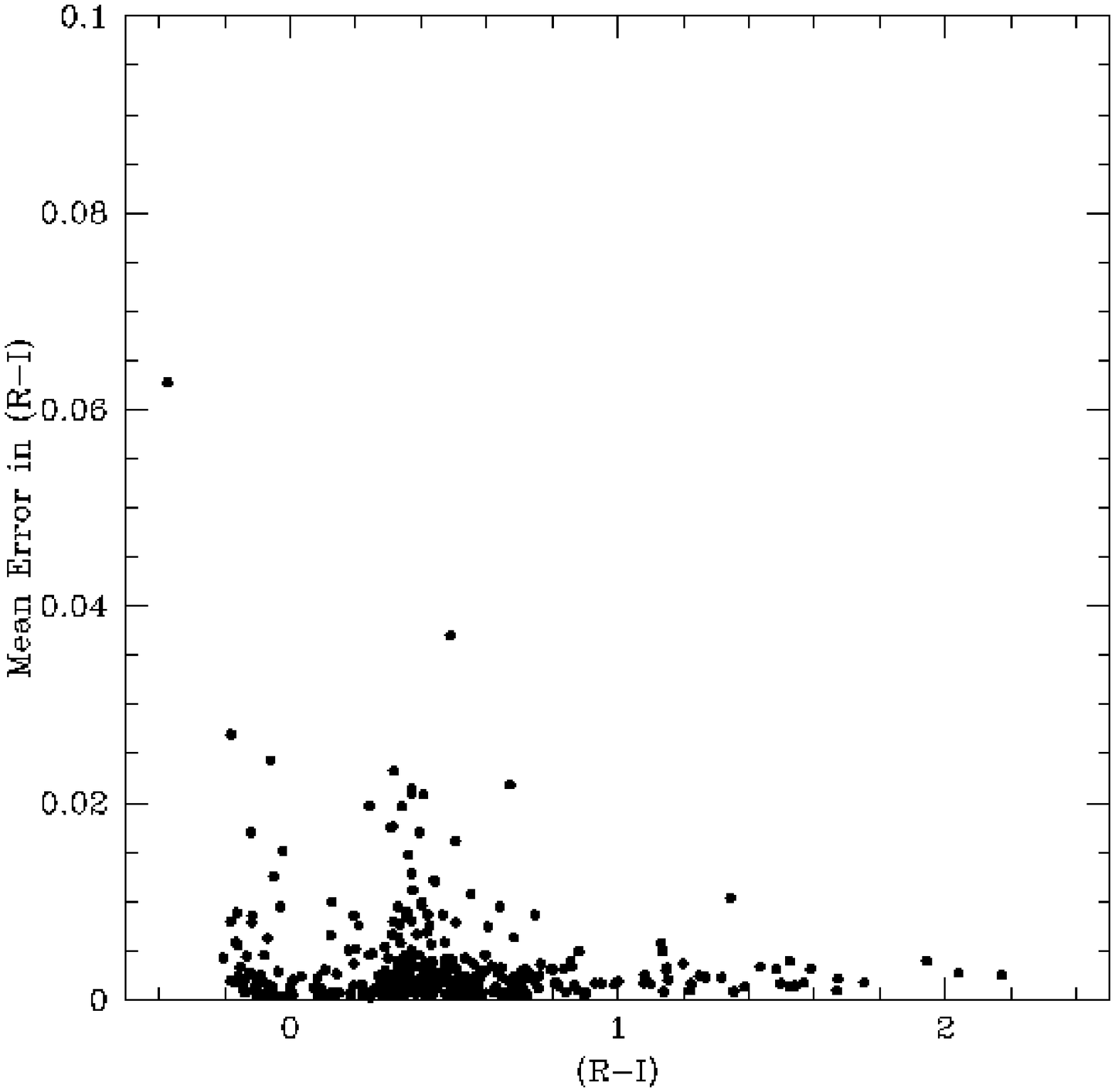}
\caption{Mean error of the mean of a single observation in $(R-I)$ for the standard stars as a function of $(R-I)$.}
\label{fig:figure144}
\end{figure}

\clearpage
\begin{figure}
\plotone{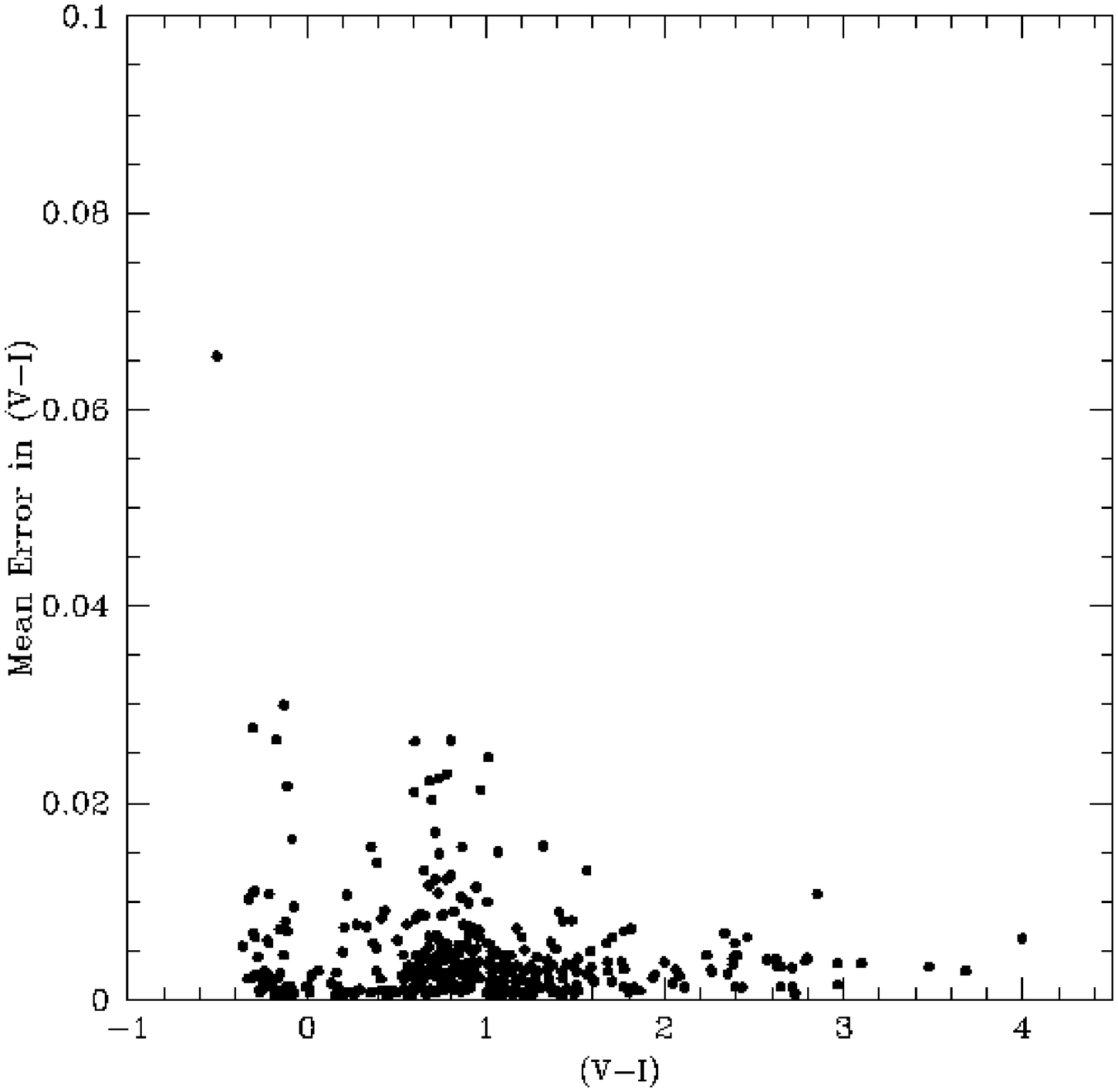}
\caption{Mean error of the mean of a single observation in $(V-I)$ for the standard stars as a function of $(V-I$).}
\label{fig:figure145}
\end{figure}

\clearpage
\begin{figure}
\plotone{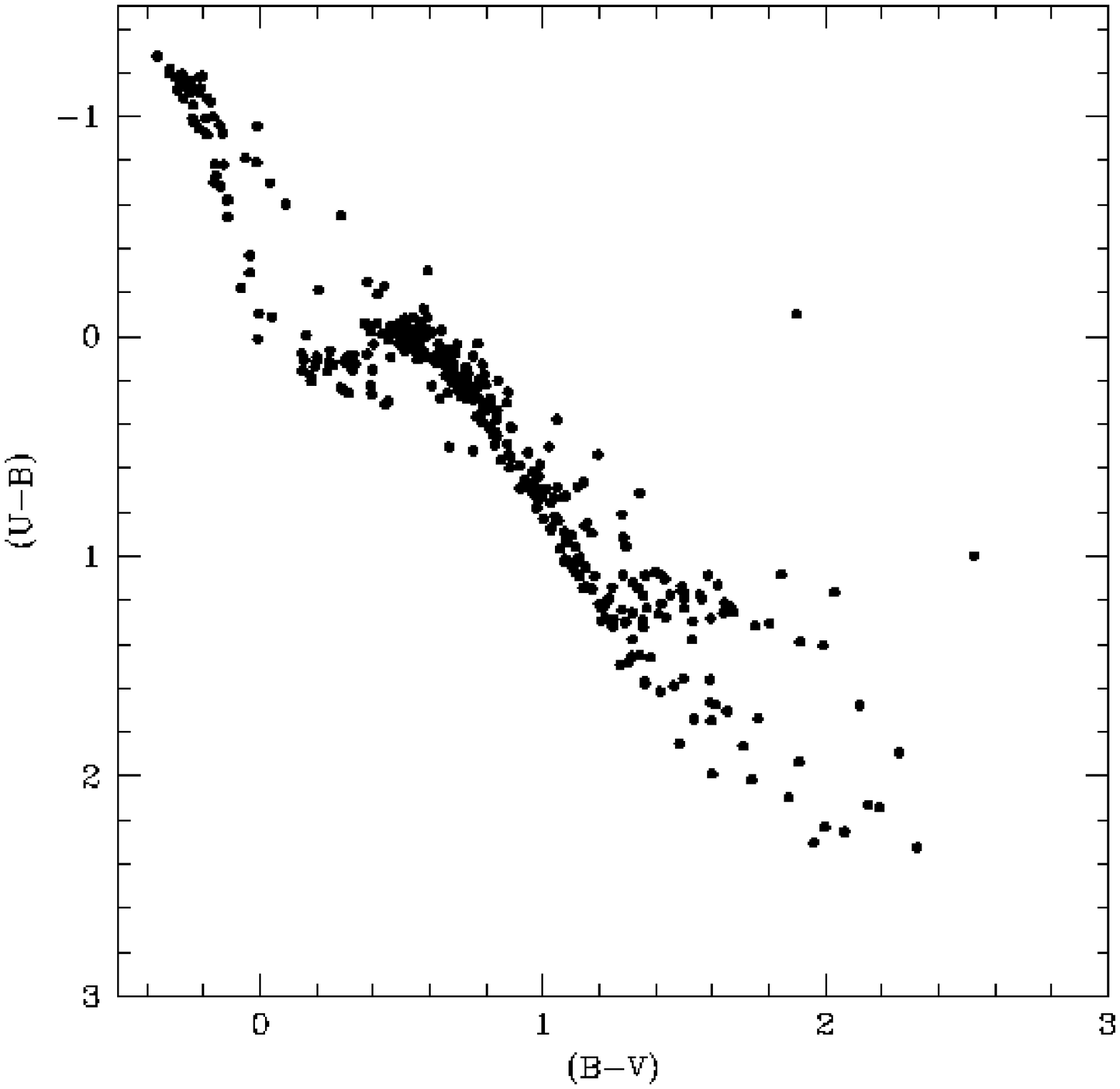}
\caption{$[(U-B), (B-V)]$ color-color plot for stars measured in this paper with five or more observations.}
\label{fig:figure146}
\end{figure}

\clearpage
\begin{figure}
\plotone{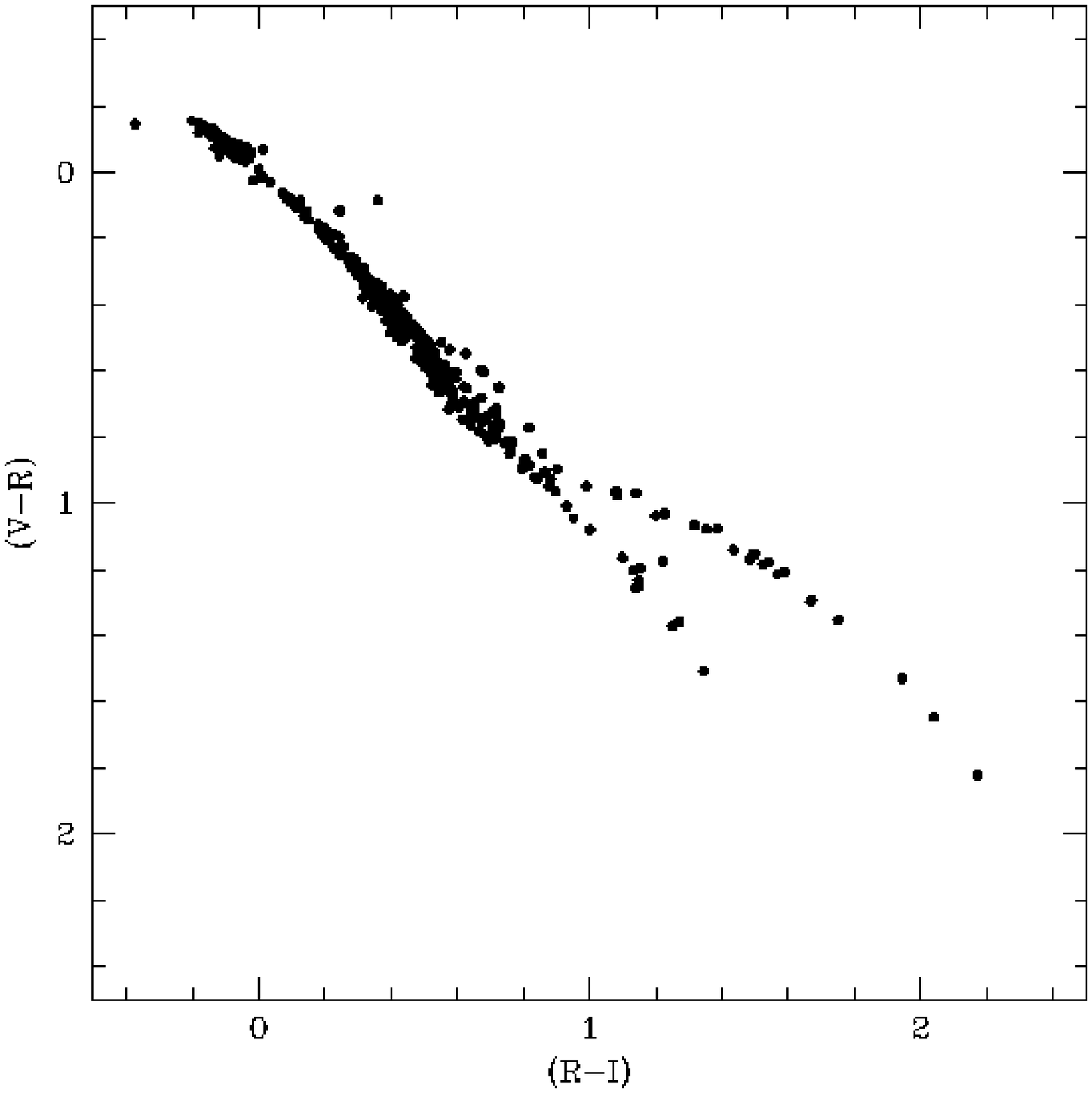}
\caption{$[(V-R), (R-I)]$ color-color plot for stars measured in this paper with five or more observations.}
\label{fig:figure147}
\end{figure}